\documentclass[structabstract]{aa}  
\usepackage{bigstrut}
\usepackage{graphicx}
\usepackage{txfonts}
\usepackage[round]{natbib}
\usepackage{url}
\usepackage{textcomp}
\def \kh{\langle\kappa\rangle_{\mathrm{H}}}
\def \kcrit{\kappa_{\mathrm{crit}}}
\def \kap{\kappa_{\mathrm{acc}}}
\def \qq{Q_{\mathrm{acc}}}
\def \ag{a_{\mathrm{gr}}}

\begin{document}
   \title{Exploring wind-driving dust species in cool luminous giants}

   \subtitle{II. Constraints from photometry of M-type AGB stars}

   \author{S. Bladh
          \inst{1}
          \and
          S. H\"ofner\inst{1}
          \and
          W. Nowotny\inst{2}
          \and
          B. Aringer\inst{2}
          \and
          K. Eriksson\inst{1}
          }
   \institute{Department of Physics and Astronomy, Division of Astronomy and Space Physics, Uppsala University,
              Box 516, SE-75120, Uppsala, Sweden\\
              \email{sara.bladh@physics.uu.se}
\and
             University of Vienna, Department of Astrophysics, T\"urkenschanzstra{\ss}e 17, A-1180 Wien, Austria
             }

   \date{Received October 18, 2012; accepted XXXX, 2012}

 
\titlerunning{Exploring wind-driving dust species in cool luminous giants II.}
\authorrunning{S. Bladh et al. } 
 
  \abstract
       {The heavy mass loss observed in evolved asymptotic giant branch (AGB) stars is usually attributed to a two-stage process: atmospheric levitation by pulsation-induced shock waves, followed by radiative acceleration of newly formed dust grains. The dust transfers momentum to the surrounding gas through collisions and thereby triggers a general outflow. Radiation-hydrodynamical models of M-type AGB stars suggest that these winds can be driven by photon scattering -- in contrast to absorption -- on Fe-free silicate grains of sizes 0.1--1\,$\mu$m.}
   {In this paper we study photometric constraints for wind-driving dust species in M-type AGB stars, as part of an ongoing effort to identify likely candidates among the grain materials observed in circumstellar envelopes.}
   {To investigate the scenario of stellar winds driven by photon scattering on dust, and to explore how different optical and chemical properties of wind-driving dust species affect photometry we focus on two sets of dynamical models atmospheres: (i) models using a detailed description for the growth of Mg$_2$SiO$_4$ grains, taking into account both scattering and absorption cross-sections when calculating the radiative acceleration, and (ii) models using a parameterized dust description, constructed to represent different chemical and optical dust properties. By comparing synthetic photometry from these two sets of models to observations of M-type AGB stars we can provide constraints on the properties of wind-driving dust species.}
     {Photometry from wind models with a detailed description for the growth of Mg$_2$SiO$_4$ grains reproduces well both the values and the time-dependent behavior of observations of M-type AGB stars, providing further support for the scenario of winds driven by photon scattering on dust. The photometry from the models with a parameterized dust description suggests that wind-drivers need to have a low absorption cross-section in the visual and near-IR to reproduce the time-dependent behavior, i.e. small variations in ($J$\,--\,$K$) and spanning a larger range in ($V$\,--\,$K$). This places constraints on the optical and chemical properties of the wind-driving dust species.}
  {To reproduce the observed photometric variations in ($V$\,--\,$K$) and ($J$\,--\,$K$) both detailed and parameterized models suggest that the wind-driving dust materials have to be quite transparent in the visual and near-IR. Consequently, strong candidates for outflows driven by photon scattering on dust grains are Mg$_2$SiO$_4$, MgSiO$_3$, and potentially SiO$_2$.}


   \maketitle

\section{Introduction}

\label{s_intro}
There is a substantial amount of observational evidence for the presence of dust in the circumstellar environment of asymptotic giant branch (AGB) stars, and it has long been argued that the slow winds detected in these stars are caused by radiative acceleration on dust particles \citep[see, e.g.,][]{wi66,ge71,sed94,agb1}. In fact, for C-type AGB stars (C/O\,$>$\,1) there is hardly any doubt that the outflows are driven by radiation pressure on carbon grains forming in the cool, extended atmospheres created by pulsation-induced shock waves. Detailed models of this scenario show good agreement with a range of observations, i.e. high resolution spectroscopy, photometry and interferometry \citep[e.g.][]{wint00,gloidl04,now10,now11,sac11}. For M-type AGB stars (C/O{\,$<$\,1), on the other hand, the construction of realistic wind models has proven more difficult \citep[e.g.][]{jeong03} and it is still a matter of debate which grain species are responsible for driving the outflows \citep[see, e.g., the discussion in][]{hof09}. Characteristic features of various dust species have been observed in the mid-IR \citep[see, e.g.,][for an overview]{dor10,agbgrain} but few of them fulfill the conditions necessary for triggering outflows: (i) able to form in the close vicinity of the star, (ii) consisting of abundant materials and (iii) large radiative cross-sections in the near-IR \citep[for a more detailed discussion see][hereafter referred to as \mbox{Paper I}]{bladh12}. 

Magnesium-iron silicates, i.e. olivine ([Mg,Fe]$_2$SiO$_4$) and pyroxene ([Mg,Fe]SiO$_3$), are commonly considered strong candidates for wind-drivers in M-type AGB stars. A high abundance of silicates in the circumstellar envelopes can be deduced from observations of characteristic features at 9.7~$\mu$m and 18~$\mu$m \citep[e.g.][]{woney69,low70,mol02a}.\footnote{The 9.7~$\mu$m  band is due to a stretching resonance in Si-O, while the 18~$\mu$m band is caused by a bending mode in the SiO$_4$ tetrahedron.} While such mid-IR features are important for identifying individual dust species it is, however, the optical dust properties in the near-IR that are critical for the wind mechanism, since most of the stellar radiation is emitted in this region. For silicate grains the absorption efficiency in the near-IR is strongly dependent on the Fe-content and silicate materials at the Mg-rich end are very transparent to near-IR radiation. Using frequency-dependent wind models with a detailed treatment of dust formation, \cite{woi06fe} demonstrated that silicate grains have to be essentially Fe-free in the close vicinity of the star; Fe-bearing silicates heat up when interacting with the radiation field and are therefore not thermally stable close to the stellar surface. The low near-IR absorption cross-sections of Fe-free grains are not sufficient to trigger outflows, which raised doubts about the scenario of dust driven winds in M-type AGB stars.

In response to the findings by \cite{woi06fe}, \citet{hof08bg} suggested that the outflows in M-type AGB stars may be driven by photon scattering on Fe-free silicates. This scenario requires grains of sizes about 0.1--1\,$\mu$m, comparable to the wavelength of the stellar flux maximum, in order for scattering to be efficient. Recently, \cite{norr12} detected dust particles of sizes $\sim0.3~\mu$m in the close circumstellar environment of three M-type AGB stars, using multi-wavelength aperture-masking polarimetric interferometry in the  near-IR. The dust grains produce a halo of scattered light around the star with a polarization tangential to the stellar surface, which can be resolved with interferometric measurements. Their results provide strong observational support that grains can grow to sizes required for triggering outflows. Further confirmation of silicate grains in the close stellar environment is provided by recent mid-IR interferometric observations of RT Vir \citep{sac13}. However, measurements at such long wavelengths are not sensitive to the grain size. Stellar winds driven by photon scattering on Fe-free silicate grains result in very low circumstellar reddening due to the transparency of this material in the near-IR wavelength region. The low degree of stellar radiation thermally reprocessed by the dusty envelope could explain why earlier dynamic models without wind \citep[e.g.][]{tej03} reproduce observed visual and near-IR spectra and photometry reasonably well.

In order to further test the scenario of stellar winds driven by photon scattering on silicate grains we here present synthetic photometry and spectra for the set of models in \citet{hof08bg} and compare them with photometric observations of M-type AGB stars. Most evolved AGB stars belong to the group of long period variables (LPVs) and the photometric magnitudes and colors change with pulsation phase. A comparison of the light variations of observed targets with the corresponding modeling results can help us test the dynamical models. To understand if the photometry resulting from these models is a trivial result, i.e. a generic property of the models, or determined specifically by the grain properties of Mg$_2$SiO$_4$, we also consider another set of models that use a parameterized dust description, constructed to represent different chemical and optical dust properties. This set of models, first presented in Paper I, allows us to investigate how different dust properties will affect the photometry and spectra.

In Paper I we focused on dynamical criteria when searching for possible wind-driving dust species in M-type AGB stars, i.e. what combination of optical and chemical properties are necessary for a dust species to be able to form close enough to the star to initiate mass outflows. Here, on the other hand, we explore the resulting spectra and photometry of the dynamical models, i.e. what optical properties the wind-driving dust species need to possess to achieve agreement with observations.

The paper is organized in the following way: in Sect. \ref{s_dynmod} we introduce the atmosphere and wind model and the two different descriptions for the dust component. In Sect. \ref{s_mpid} we present the parameters used in the models of \cite{hof08bg} and the models included from Paper I, respectively, and also wind properties of individual models. A description concerning the details of the spectral synthesis is given in Sect. \ref{s_modspe}, the observational data sets are presented in Sect.~\ref{s_obsdat} and the photometric results from the different model sets are given in Sects.~\ref{s_phdetdus}--\ref{s_phpardus}. We comment on specific dust species in Sect. \ref{s_spg} and in Sect. \ref{s_concl2} we provide a summary of our conclusions.

\section{Modeling of atmosphere and wind}
\label{s_dynmod}

The dynamical models cover a spherical shell with an inner boundary situated just below the photosphere and an outer boundary located at the point where the wind velocity has reached its terminal value, i.e. around 20--30\,R$_*$.\footnote{In this context the stellar radius is defined by $R_*=\sqrt{L_\star/4\pi\sigma T_\star^4}$, where $L_\star$ is the luminosity and $T_\star$ is the effective temperature of the star. The inner boundary is situated typically a few percent below this point.} The variable structures of the atmospheres are described by the equations of hydrodynamics (equation of continuity, equation of motion and energy equation) and the pulsations are simulated by temporal variation of physical quantities at the inner boundary. The opacities of molecules and dust which form in the outer cool layers of the atmospheres dominate the radiation field, and in order to achieve realistic density--temperature structures the models include a frequency-dependent treatment of the radiative transfer \citep[see][and Paper I for more details]{hof03,hof08bg}. The models feature two different descriptions for the dust component: (i) a time-dependent description for the growth of Mg$_2$SiO$_4$ grains with a grain-size dependent treatment of the optical properties (see Sect.~\ref{s_detdus}) or (ii) a parameterized dust description based on a simplified treatment of both the grain growth and the optical properties (see Sect.~\ref{s_pardus}). In both cases the detailed dynamical models produce snapshots of the radial structure of the atmosphere, and thereby provide information about properties such as velocity, temperature, density and degree of condensation for the dust component as a function of radial distance and time. 

\subsection{Detailed dust description (D models)}
\label{s_detdus}

\begin{figure}
\centering
\includegraphics[width=9cm]{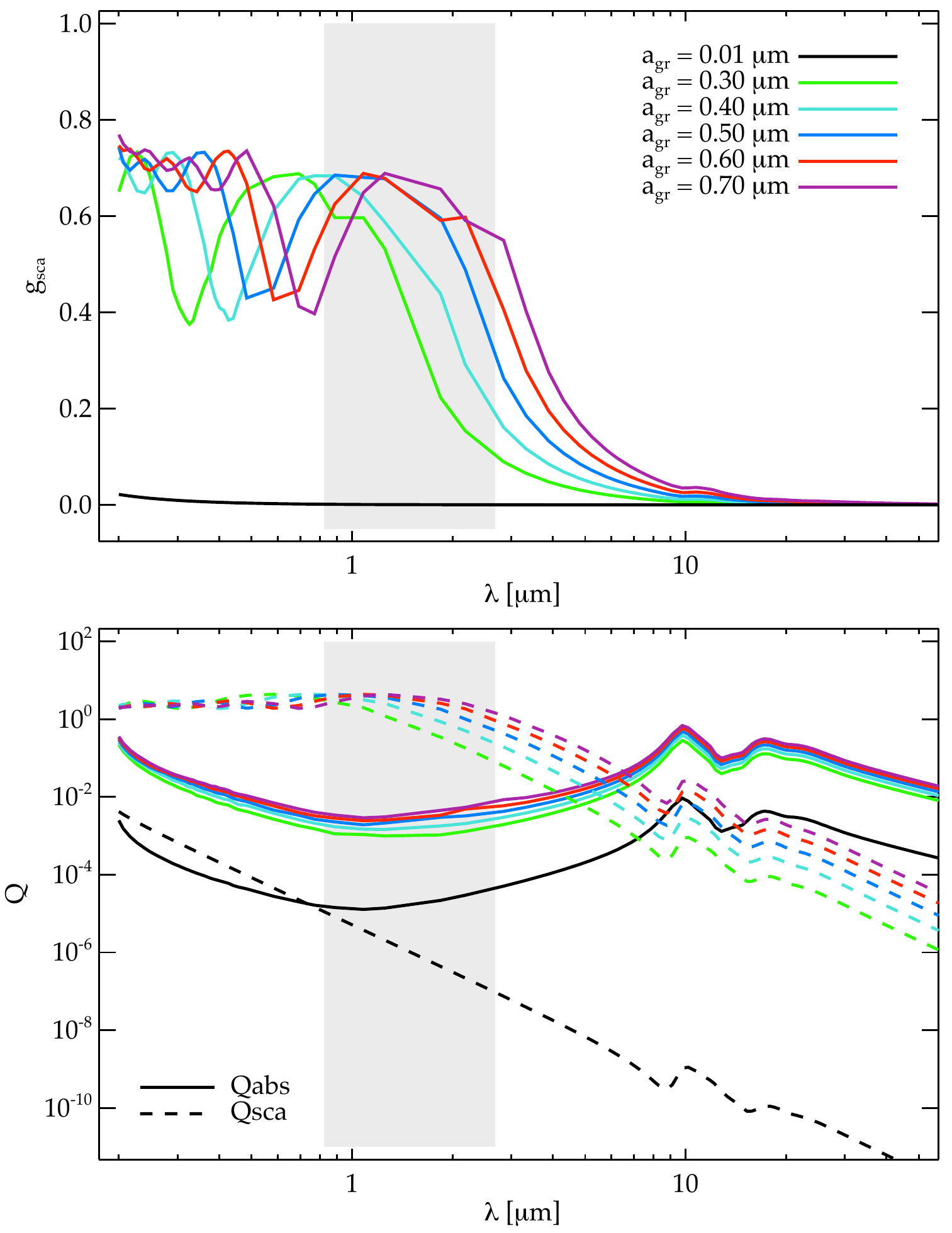}
   \caption{Optical properties of spherical Mg$_2$SiO$_4$ grains as a function of wavelength, calculated for grain radii varying between $0.3-0.7\,\mu$m, as well as grains in the small particle limit ($a_{\mathrm{gr}}=0.01\,\mu$m, black curves). These properties were calculated from optical data by \citet{jag03}, using Mie theory. The shaded area marks the region where most of the stellar flux is emitted.  \textit{Top panel:} The asymmetry factor $g_{\mathrm{sca}}$. \textit{Bottom panel:} The absorption and scattering efficiencies, $Q_{\mathrm{abs}}$ and $Q_{\mathrm{sca}}$, here smoothed to avoid unrealistic resonances inherent to Mie theory.}
    \label{f_gsca}
\end{figure}

In the models presented by \cite{hof08bg}, the growth of pure forsterite particles (Mg$_2$SiO$_4$) is modeled according to the net reaction
\begin{equation}
2\mathrm{Mg }+\mathrm{SiO}+3\mathrm{H}_2\mathrm{O} \longrightarrow \mathrm{Mg}_2\mathrm{SiO}_4+3\mathrm{H}_2,
\end{equation}
under the assumption that the step determining the total growth rate is the addition of SiO molecules to the grain surface. The equation describing the growth and decomposition of Mg$_2$SiO$_4$ grains follows \cite{gail99} and is given by\footnote{Note that this equation is given here in the co-moving frame, in contrast to the differential equations of radiation-hydrodynamics in \mbox{Paper I.}}
\begin{eqnarray}
\label{e_gr}
\frac{da_{\mathrm{gr}}}{dt} &=& V_{\mathrm{Mg}_2\mathrm{SiO}_4}\left[J^{\mathrm{gr}}_{\mathrm{SiO}}-J^{\mathrm{dec}}_{\mathrm{SiO}}\right]\nonumber \\
&=&V_{\mathrm{Mg}_2\mathrm{SiO}_4}\alpha_{\mathrm{SiO}}\textit{v}_{\mathrm{SiO}}n_{\mathrm{SiO}}\left[1-\frac{p_{\mathrm{v,SiO}}}{n_{\mathrm{SiO}}\,kT_{\mathrm{g}}}\sqrt{\frac{T_{\mathrm{g}}}{T_{\mathrm{d}}}}\right],
\end{eqnarray}
where $a_{\mathrm{gr}}$ is the grain radius, assuming spherical grains. $J^{\mathrm{gr}}_{\mathrm{SiO}}$ and $J^{\mathrm{dec}}_{\mathrm{SiO}}$ denote the growth and decomposition rate of SiO molecules per grain surface area, $V_{\mathrm{Mg}_2\mathrm{SiO}_4}$ is the volume of the monomer (the basic building block of the grain material), $\alpha_{\mathrm{SiO}}$ is a sticking coefficient, $\textit{v}_{\mathrm{SiO}}$ is the thermal velocity of the SiO molecules, $n_{\mathrm{SiO}}$ is the number density of SiO molecules in the gas and $p_{\mathrm{v,SiO}}$ is the hypothetical partial pressure of SiO molecules in chemical equilibrium between the gas phase and the solid. For the atmospheres of M-type AGB stars there is currently no well-established nucleation theory. For simplicity we assume the existence of seed particles that start to grow when the atmospheric environment favors grain growth. This will result in a uniform grain size for all dust particles at a given distance from the stellar surface. 

The output of the equation describing the grain growth (Eq.~(\ref{e_gr})) is the grain radius $a_{\mathrm{gr}}$ at a given distance and time. If Mg$_2$SiO$_4$ particles grow to grain radii comparable to the wavelength of the flux maximum, the contribution to the radiative acceleration from the scattering cross-section is substantial, dominating over true absorption by several orders of magnitude, as can be seen in the lower panel of Fig.~\ref{f_gsca}. To include the contribution from photon scattering in the radiative acceleration, i.e. in the opacity entering the equation of motion (see Sect. 3.1 Paper I), we calculate the grain-size dependent dust opacity per mass accordingly
\begin{equation}
\label{eq17}
\kappa_{\mathrm{acc}}(\lambda,a_{\mathrm{gr}})=\frac{3}{4}\frac{A_{\mathrm{mon}}}{\rho_{\mathrm{gr}}}\frac{\qq(\lambda,\ag)}{\ag}\frac{\varepsilon_\mathrm{Si}}{1+4\varepsilon_{\mathrm{He}}} f_\mathrm{c},
\end{equation}
see Sect. 2.1 in Paper I for details. Here $A_{\mathrm{mon}}$ is the atomic weight of the monomer,  $\rho_{\mathrm{gr}}$ the bulk density of the grain material, $\varepsilon_{\mathrm{Si}}$ and $\varepsilon_{\mathrm{He}}$ the abundances of silicon and helium, respectively, and $f_\mathrm{{c}}$ the degree of condensation. The efficiency $\qq(\lambda,a_{\mathrm{gr}})$, i.e. the radiative cross-section divided by the geometrical cross-section of the grain, is defined by
\begin{equation}
\label{e_qtot}
\qq= Q_{\mathrm{abs}} + (1-g_{\mathrm{sca}})Q_{\mathrm{sca}}
\end{equation}
where $Q_{\mathrm{abs}}$ and $Q_{\mathrm{sca}}$ are the efficiencies of absorption and scattering, respectively, and $g_{\mathrm{sca}}$ is the asymmetry factor describing deviations from isotropic scattering (a value of zero indicates isotropic scattering and a value of one pure forward scattering). These quantities can be computed from optical data, using Mie theory \citep[program BHMIE\footnote{\url{http://www.astro.princeton.edu/~draine/scattering.html}} from][modified by Draine]{bohuff83}. Figure~\ref{f_gsca} shows the absorption and scattering efficiencies, as well as the asymmetry factor, as a function of wavelength for forsterite grains of different sizes. As can be seen, grains in the size range $0.3-0.7\,\mu$m (relevant for the models in set D) lead to predominantly forward scattering in the wavelength region where most of the stellar flux is emitted. This means that only a fraction of the scattering events will transfer momentum to the dust particles, which is taken into account by the factor $(1-g_{\mathrm{sca}})$ in Eq.~\ref{e_qtot}, and consequently in the equation of motion.

The degree of condensation, i.e., the fraction of silicon bound in dust compared to the total amount of silicon, can be computed from the grain radius $a_{\mathrm{gr}}$ (given by (Eq.~(\ref{e_gr})), the volume of the monomer $V_{\mathrm{Mg}_2\mathrm{SiO}_4}$, the abundance of seed particles $n_{\mathrm{gr}}/n_{\mathrm{H}}$ and the elemental abundance of silicon ${\varepsilon_{\mathrm{Si}}}$,
\begin{equation}
\label{eq18}
f_\mathrm{c}(r,t)=\frac{4\pi a_{\mathrm{gr}}^3(r,t)}{3}\frac{1}{V_{\mathrm{Mg}_2\mathrm{SiO}_4}}\frac{n_{\mathrm{gr}}}{n_{\mathrm{H}}}\frac{1}{\varepsilon_{\mathrm{Si}}}.
\end{equation}
The only free input parameter in this model for grain growth is the abundance of seed particles.\footnote{This quantity is defined here as the number of grains per hydrogen nucleus, expressed as the ratio of the number density of grains to the total number of hydrogen particles per volume of atmosphere.} Generally, within the range where observable wind properties are still reproduced, a lower abundance of seed particles will result in condensates with larger grain radius, due to less competition in accumulating the surrounding material, whereas a higher abundance of seed particles will produce smaller grains.

\subsection{Parameterized dust description (P models)}
\label{s_pardus}

\begin{figure}
\centering
\includegraphics[width=8.5cm]{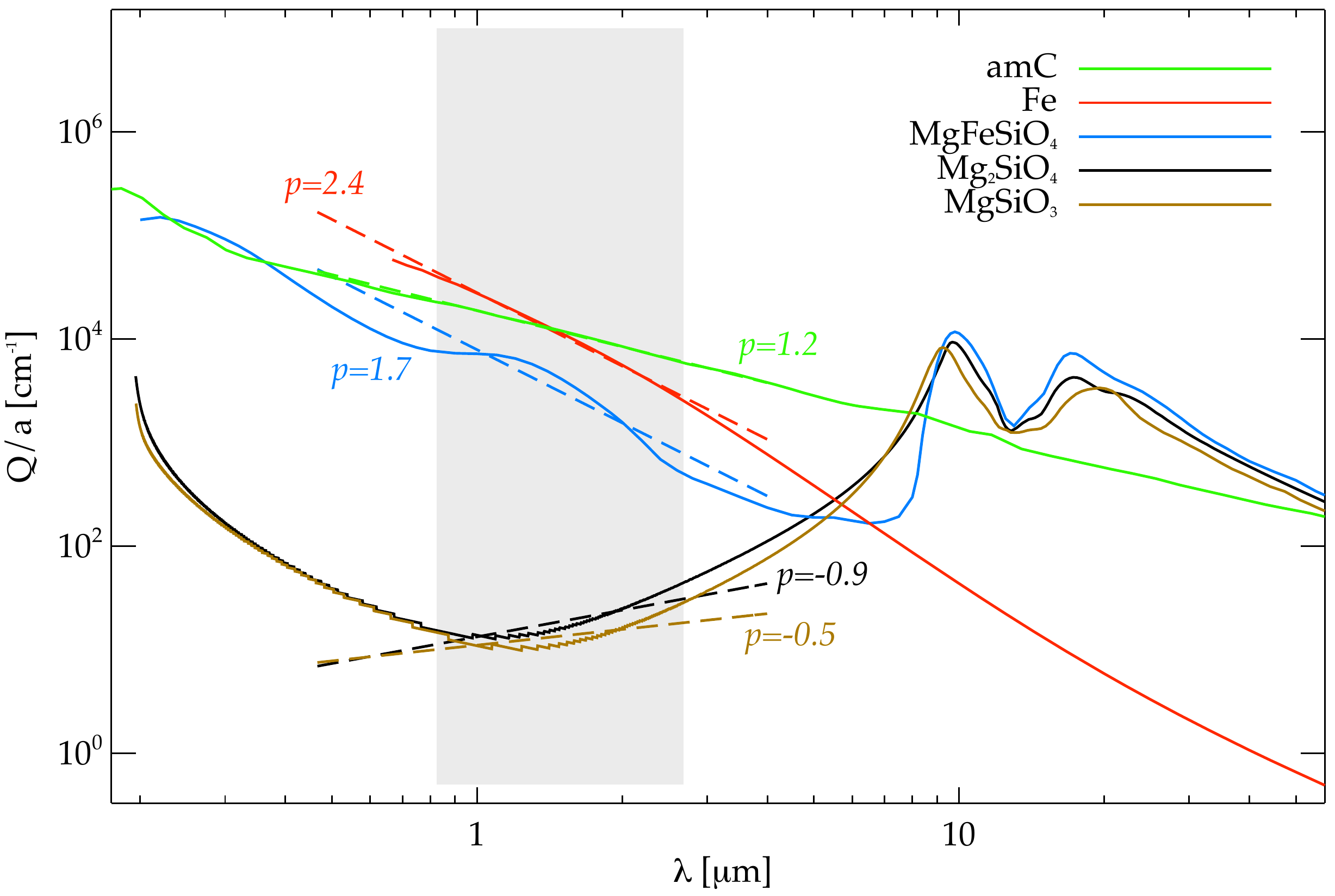}
   \caption{Efficiency per grain radius $\qq/\ag$ in the small particle limit, as a function of wavelength, for a selection of dust species. The dashed lines show the power law fit corresponding to Eq. (\ref{e_pl}) and the shaded area indicates the wavelength region for which the optical data is fitted (for references to the sources of optical data see Table~\ref{t_more}).}
    \label{f_pl}
\end{figure}
In Paper I we introduced a parameterized dust description in order to systematically study the effects of a range of optical and chemical dust properties on the dynamics and the spectral energy distribution of the model atmospheres. The formula uses a simplified description of grain growth, expressed in terms of $f_{\mathrm{c}}$, in combination with a wavelength-dependent power-law function $\hat{\kappa}(\lambda)$ describing different optical properties
\begin{equation}
\label{e_parkap1}
\kap(\lambda) = \hat{\kappa}(\lambda)\cdot f_{\mathrm{c}}(r,t,T_{\mathrm{c}}).
\end{equation}
Our description was inspired by a formula used by \cite{b88} in his dust driven wind models, but is generalized to allow for wavelength-dependent optical properties. The degree of condensation $f_{\mathrm{c}}$ describing the grain growth is designed to increase monotonically with falling grain temperature, approaching unity as the grain temperature $T_{\mathrm{d}}$ drops well below the condensation temperature $T_{\mathrm{c}}$ of the condensate.
\begin{equation}
\label{e_fc}
f_{\mathrm{c}}(r,t,T_{\mathrm{c}}) = \frac{1}{1+e^{(T_{\mathrm{d}}(r,t)-T_{\mathrm{c}})/\Delta T}}
\end{equation}
Given the low density of the circumstellar environment and the proximity to a strong radiation source, the grain temperature is assumed to be determined by radiative equilibrium rather than collisions with gas particles. Because of the varying radiation field the grain temperature is a function of both time and distance from the star. The parameter $\Delta T$ regulates the width of the dust formation zone (see Fig.~3 of Paper I).

The optical properties of the dust material are modeled by a power-law function
\begin{equation}
 \label{e_pl}
\hat{\kappa}(\lambda) = \kappa_{\mathrm{0}}\left(\frac{\lambda}{\lambda_0}\right)^{-p},
\end{equation}
where $p$ is obtained by fitting a power-law to the efficiency per grain radius $\qq/\ag$, assuming small particles, in the wavelength region where most of the stellar flux is emitted (see the shaded area in Fig. \ref{f_pl}). In this expression $\kappa_{\mathrm{0}}$ is a scaling factor such that $\hat{\kappa}(\lambda_0)=\kappa_0$. 

To distinguish between effects of scattering and true absorption of photons on dust grains we further introduce a quantity $f_{\mathrm{abs}}$ which sets the percentage of the dust opacity $\kap$ that is to be considered as true absorption
\begin{equation}
\label{e_parkap2}
\kappa_{\mathrm{abs}}(\lambda) = f_{\mathrm{abs}}\cdot\hat{\kappa}(\lambda)\cdot f_c(r,t)\quad\mathrm{where}\quad f_{\mathrm{abs}}=\frac{\kappa_{\mathrm{abs}}}{\kap}
\end{equation}
The dust opacity $\kap$, which includes contributions from both true absorption and scattering, is used to calculate the radiative acceleration in the equation of motion, and the true absorption part $\kappa_{\mathrm{abs}}$ is used to determine the grain temperature. This allows us to separate the dynamical and thermal effects of the dust opacity, and the parameter $f_{\mathrm{abs}}$ can be adjusted to explore the effects of varying degrees of true absorption.

\section{Dynamical models: parameters and properties}
\label{s_mpid}

\subsection{Set D  --  \textbf{D}etailed dust description}
\label{s_mdetdust}

\begin{table*}
\caption{Model parameters, dynamical properties and dust characteristics for the models in set D.}             
\label{t_parmodbg}      
\centering                          
\begin{tabular}{c c c c c c c c c c c}        
\hline\hline                 
DMA & $M_\star$ & $L_\star$ & $T_\star$ & $P$ & $\Delta u_{p}$ & $n_{\mathrm{gr}}/n_{\mathrm{H}}$ & $\langle \dot{M} \rangle$ & $\langle u \rangle$ & $\langle f_{\mathrm{Si}} \rangle$ & $\langle a_{\mathrm{gr}} \rangle$ \bigstrut[t]\\  
 & $[M_{\odot}$] & [$L_{\odot}$]  & [K] &  [d] & [km/s] & & [$M_{\odot}$/yr] & [km/s] &  & [$\mu$m] \bigstrut[b]\\    
\hline       
 A2 & 1 & 5000  & 2800 & 310 & 4.0 & $1\times 10^{-15}$& $4\times10^{-7}$ & 5 & 0.18 & 0.48 \bigstrut[t]\\
 A3 & 1 & 5000  & 2800 & 310 & 4.0 & $3\times 10^{-15}$& $8\times10^{-7}$ & 10 & 0.22 & 0.36\\
 B1 & 1 & 7000  & 2700 & 390 & 4.0 & $3\times 10^{-16}$& $5\times10^{-7}$ & 4 & 0.14 & 0.66\\
 B2 & 1 & 7000  & 2700 & 390 & 4.0 & $1\times 10^{-15}$& $8\times10^{-7}$ & 7 & 0.15 & 0.45\\
 B3 & 1 & 7000  & 2700 & 390 & 4.0 & $3\times 10^{-15}$& $1\times10^{-6}$ & 11 & 0.20 & 0.34\\
 C1 & 1 & 10000  & 2600 & 525 & 4.0 & $3\times 10^{-16}$& $3\times10^{-6}$ & 7 & 0.13 & 0.62 \bigstrut[b]\\
 \hline
\end{tabular}
\tablefoot{Columns 2--7 list the input parameters of the models: mass  $M_\star$, luminosity $L_\star$, effective temperature $T_\star$, period P and piston velocity amplitude $\Delta u_{p}$ and the assumed seed particle abundance $n_{\mathrm{gr}}/n_{\mathrm{H}}$. Columns 8--9 list the resulting wind and dust properties properties: mass-loss rate $\langle \dot{M}\rangle$, terminal wind velocity $\langle u \rangle$, degree of condensation $\langle f_{\mathrm{Si}} \rangle$ of the key element Si and grain radius $\langle a_{\mathrm{gr}}\rangle$. Note that the resulting wind and dust properties are temporal means as indicated by the angular brackets. The optical data ($n$, $k$) for Mg$_2$SiO$_4$ are taken from \cite{jag03}.}
\end{table*}

\begin{figure}
\centering
\includegraphics[width=8.5cm]{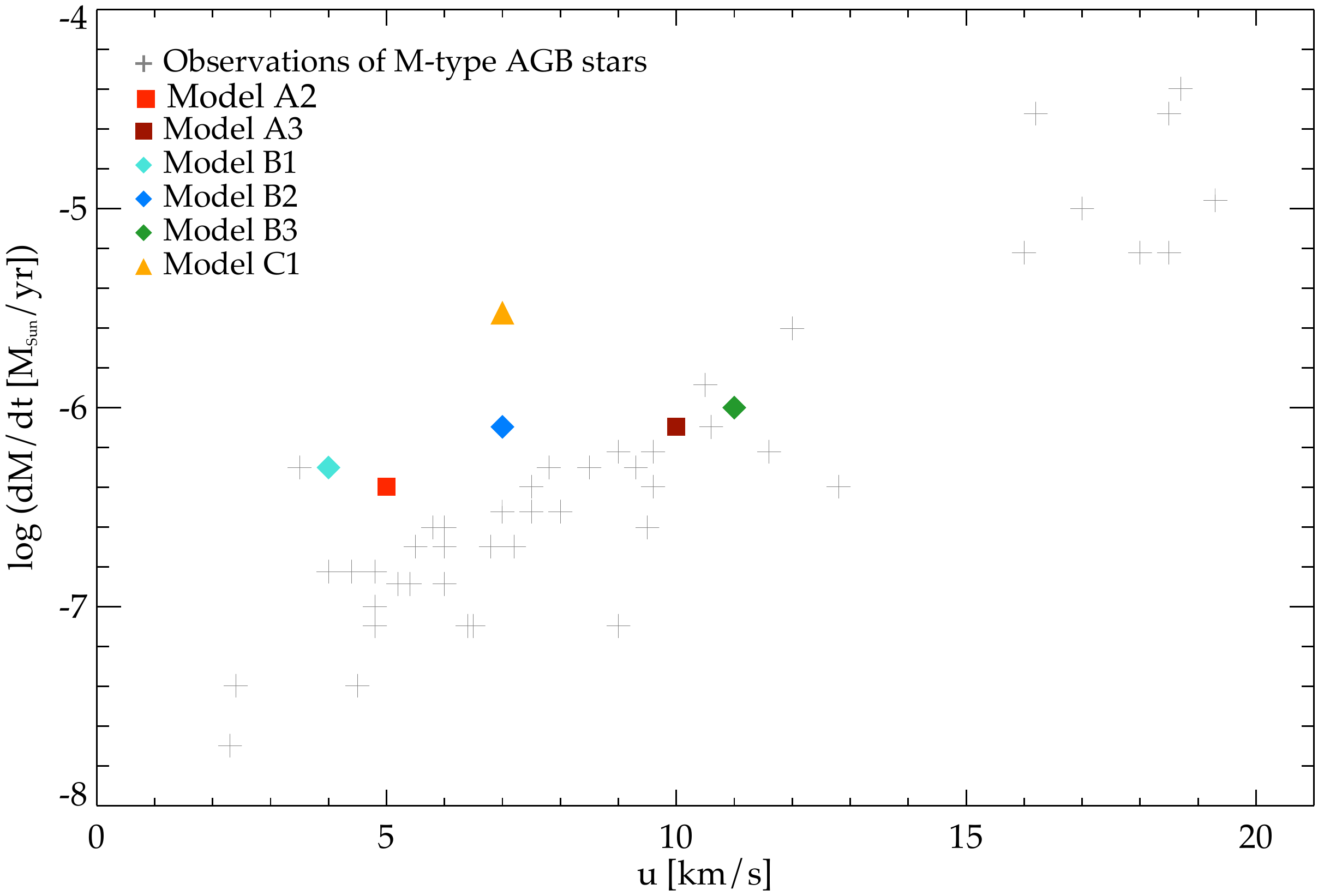}
\includegraphics[width=8.5cm]{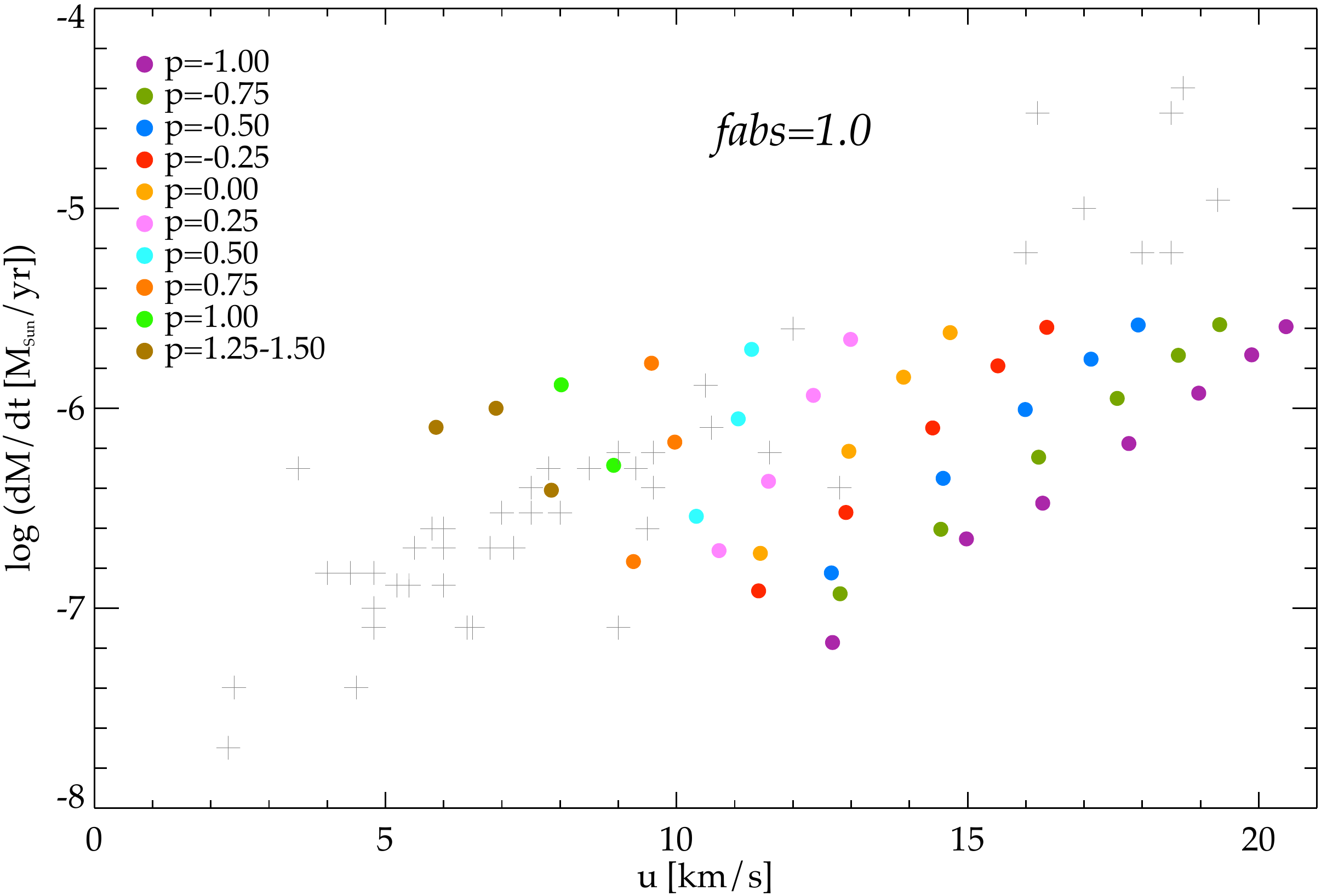}
\includegraphics[width=8.5cm]{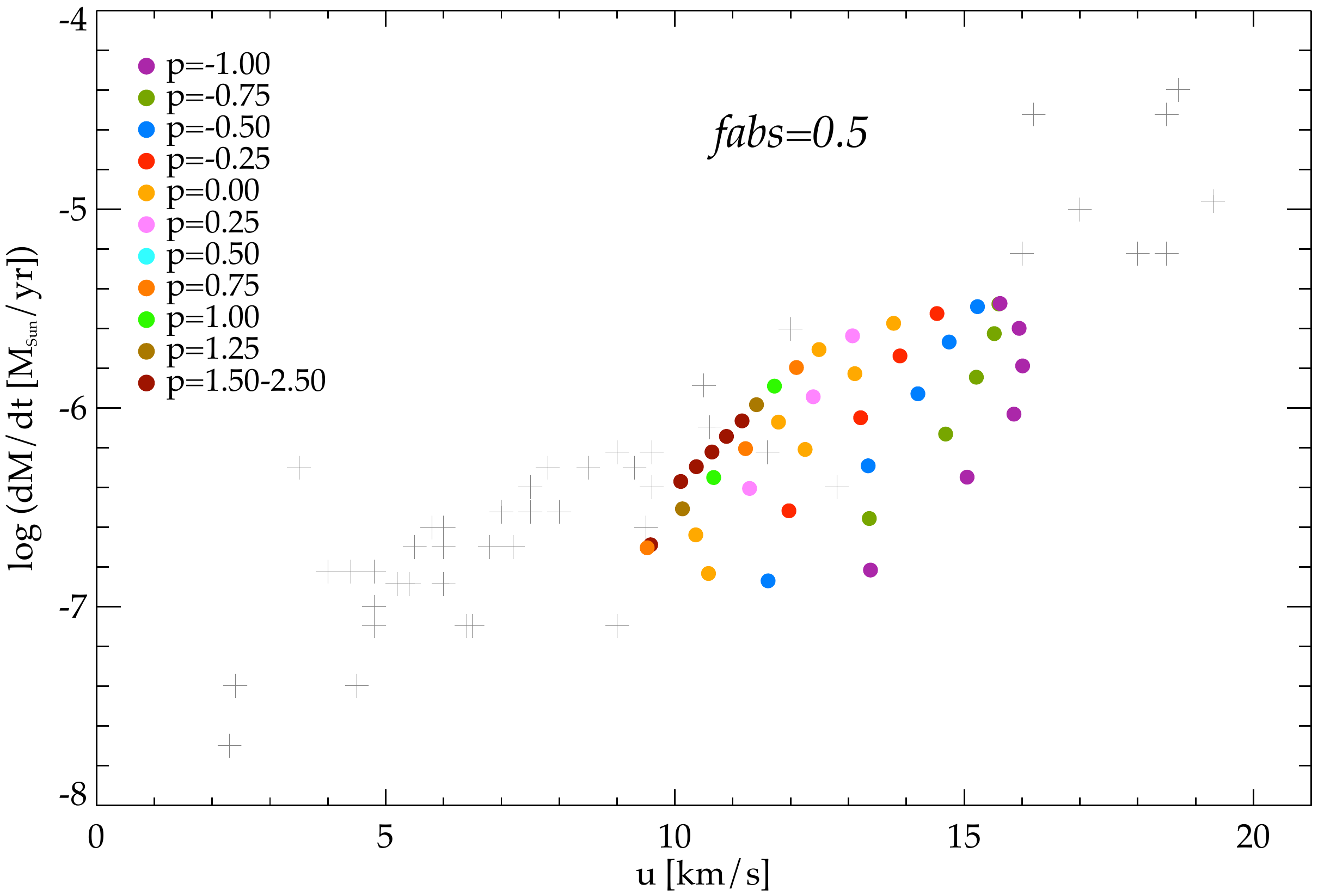}
   \caption{Observed mass-loss rates vs. wind velocities of M-type AGB stars \citep[][plus signs]{hans02,gondel03} and the corresponding properties for the dynamical models (filled circles). \textit{Top panel:} dynamical properties of the models in set D. \textit{Middle and bottom panel:} dynamical properties of the models in set P, color-coded according to $p$-value. Generally, within a set of models with same $p$ (e.g. the dark green sequence with $p=-0.75$) the mass-loss rate increases with increasing condensation temperature. For a more detailed discussion on the dynamical properties of model set P, see Paper I.}
    \label{f_dynobs}
\end{figure}

This set of dynamical model atmospheres (DMAs) was first presented in \cite{hof08bg} and is here expanded to include a model with more extreme stellar parameters in order to investigate the the effects of a higher mass-loss rate (model C1 in Tab~\ref{t_parmodbg}). The models include a detailed description for the growth of Mg$_2$SiO$_4$ grains (see Sect.~\ref{s_detdus}) and take into account the effects of photon scattering on dust particles when calculating the radiative acceleration. The outflows produced in these models are, in fact, driven by scattering of stellar photons on grains with radii between 0.1--1.0 $\mu$m, not by true absorption (which is negligible in this context). 

The stellar parameters for the different models (denoted by the letters A--C) are listed in columns 2--4 in Table~\ref{t_parmodbg}. The period in column 5 follows the period--luminosity relation from \cite{fei89}. The piston velocity amplitude (column 6) used to simulate pulsations at the inner boundary was set to 4 km/s, a reasonable value when comparing with observations of molecular lines forming in the inner atmospheric region \mbox{\citep{now10}}. The only free input parameter in the description for grain growth, the abundance of seed particles  $n_{\mathrm{gr}}/n_{\mathrm{H}}$ (denoted by the number in the model name) is listed in column 7. The resulting dynamical properties and dust characteristics are listed in columns 8--11. 

Observed mass-loss rates versus wind velocities of M-type AGB stars \citep[derived from profiles of CO radio lines,][]{hans02,gondel03} and the corresponding properties of the models in set D are shown in the top panel of Fig.~\ref{f_dynobs}. Note that the models with stellar parameters according to combinations A and B, already presented in \cite{hof08bg}, agree reasonably well with observations, whereas model C1, added here for studying the effect of a higher mass-loss rate, is further away from the region covered by the observational sample.

\subsection{Set P  --  \textbf{P}arameterized dust description}
\begin{table}
\caption{Dynamical properties of the models in set P with parameterized dust opacity. The table shows mass-loss rate $\langle \dot{M} \rangle$ and terminal wind velocity $\langle u \rangle$, in addition to the combination of $p$ and $T_{\mathrm{c}}$ used in the parameterized dust description.}             
\label{t_dynmod}      
\centering                          
\begin{tabular}{r c c c c c c}        
\hline\hline                 
& &  &   \multicolumn{2}{c}{$f_{\mathrm{abs}}=1.0$} & \multicolumn{2}{c}{$f_{\mathrm{abs}}=0.5$} \bigstrut[t] \bigstrut[b]\\    
\hline
DMA & $p$ &  $T_{\mathrm{c}}$ & $\langle \dot{M} \rangle$ & $\langle u \rangle$ & $\langle \dot{M} \rangle$ & $\langle u \rangle$ \bigstrut[t] \\    
& & [K]  &  [$M_{\odot}$/yr] & [km/s] &  [$M_{\odot}$/yr] &  [km/s] \bigstrut[b] \\   
\hline
P01 & -1.00 & 1100  & $7\times 10^{-8}$ & 13  & -                 & -    \bigstrut[t]\\
P02 & -1.00 & 1200  & $2\times 10^{-7}$ & 15  & $2\times 10^{-7}$ & 14 \\
P03 &-1.00 & 1300  & $3\times 10^{-7}$ & 16  & $5\times 10^{-7}$ & 15 \\
P04 &-1.00 & 1400  & $7\times 10^{-7}$ & 18  & $9\times 10^{-7}$ & 16 \\
P05 &-1.00 & 1500  & $1\times 10^{-6}$ & 19  & $2\times 10^{-6}$ & 16 \\
P06 &-1.00 & 1600  & $2\times 10^{-6}$ & 20  & $3\times 10^{-6}$ & 16 \\
P07 &-1.00 & 1700  & $3\times 10^{-6}$ & 20  & $3\times 10^{-6}$ & 16 \bigstrut[b]\\
\hline       
P08 &-0.75 & 1200  & $1\times 10^{-7}$ & 13 & -                 & -     \bigstrut[t]\\
P09 &-0.75 & 1300  & $2\times 10^{-7}$ & 15  & $3\times 10^{-7}$ & 13  \\
P10 &-0.75 & 1400  & $6\times 10^{-7}$ & 16  & $7\times 10^{-7}$ & 15  \\
P11 &-0.75 & 1500  & $1\times 10^{-6}$ & 18  & $1\times 10^{-6}$ & 15  \\
P12 &-0.75 & 1600  & $2\times 10^{-6}$ & 19  & $2\times 10^{-6}$ & 16  \\
P13 &-0.75 & 1700  & $3\times 10^{-6}$ & 19  & $3\times 10^{-6}$ & 16  \bigstrut[b]\\
\hline  
P14 &-0.50 & 1300  & $2\times 10^{-7}$ & 13  & $2\times 10^{-7}$ & 12  \bigstrut[t]\\
P15 &-0.50 & 1400  & $4\times 10^{-7}$ & 15  & $5\times 10^{-7}$ & 13  \\
P16 &-0.50 & 1500  & $1\times 10^{-6}$ & 16  & $1\times 10^{-6}$ & 14  \\
P17 &-0.50 & 1600  & $2\times 10^{-6}$ & 17  & $2\times 10^{-6}$ & 15  \\
P18 &-0.50 & 1700  & $3\times 10^{-6}$ & 18  & $3\times 10^{-6}$ & 15  \bigstrut[b]\\
\hline  
P19 &-0.25 & 1300  & $1\times 10^{-7}$ & 11  & -                 & -     \bigstrut[t]\\
P20 &-0.25 & 1400  & $3\times 10^{-7}$ & 13  & $3\times 10^{-7}$ & 12  \\
P21 &-0.25 & 1500  & $8\times 10^{-7}$ & 14  & $9\times 10^{-7}$ & 13  \\
P22 &-0.25 & 1600  & $2\times 10^{-6}$ & 16  & $2\times 10^{-6}$ & 14  \\
P23 &-0.25 & 1700  & $3\times 10^{-6}$ & 16  & $3\times 10^{-6}$ & 15  \bigstrut[b]\\
\hline  
P24 &0.00  & 1400  & $2\times 10^{-7}$ & 11  & $2\times 10^{-7}$ & 11 \bigstrut[t]\\
P25 &0.00  & 1500  & $6\times 10^{-7}$ & 13  & $6\times 10^{-7}$ & 12 \\
P26 &0.00  & 1600  & $1\times 10^{-6}$ & 14  & $2\times 10^{-6}$ & 13 \\
P27 &0.00  & 1700  & $2\times 10^{-6}$ & 15  & $3\times 10^{-6}$ & 14 \bigstrut[b]\\
\hline  
P28 &0.25  & 1400  & $2\times 10^{-7}$ & 11  & -                 & -      \bigstrut[t]\\
P29 &0.25  & 1500  & $4\times 10^{-7}$ & 12  & $4\times 10^{-7}$ & 11   \\
P30 &0.25  & 1600  & $1\times 10^{-6}$ & 12  & $1\times 10^{-6}$ & 12   \\
P31 &0.25  & 1700  & $2\times 10^{-6}$ & 13  & $2\times 10^{-6}$ & 13   \bigstrut[b]\\
\hline  
P32 &0.50  & 1500  & $3\times 10^{-7}$ & 10  & $2\times 10^{-7}$ & 10  \bigstrut[t]\\
P33 &0.50  & 1600  & $9\times 10^{-7}$ & 11  & $9\times 10^{-7}$ & 12  \\
P34 &0.50  & 1700  & $2\times 10^{-6}$ & 11  & $2\times 10^{-6}$ & 12 \bigstrut[b]\\
\hline  
P35 &0.75  & 1500  & $2\times 10^{-7}$ & 9    & $2\times 10^{-7}$ & 10   \bigstrut[t]\\
P36 &0.75  & 1600  & $7\times 10^{-7}$ & 10  & $6\times 10^{-7}$ & 11   \\
P37 &0.75  & 1700  & $2\times 10^{-6}$ & 10  & $2\times 10^{-6}$ & 12  \bigstrut[b]\\
\hline  
P38 &1.00  & 1600  & $5\times 10^{-7}$ & 9   & $5\times 10^{-7}$ & 11  \bigstrut[t]\\
P39 &1.00  & 1700  & $1\times 10^{-6}$ & 8   & $1\times 10^{-6}$ & 12  \bigstrut[b]\\
\hline  
P40 &1.25  & 1600  & $4\times 10^{-7}$ & 8   & $3\times 10^{-7}$ & 10  \bigstrut[t]\\
P41 &1.25  & 1700  & $1\times 10^{-6}$ & 7   & $1\times 10^{-6}$ & 11  \bigstrut[b]\\
\hline
P42 &1.50  & 1600  & -                 & -     & $4\times 10^{-7}$ & 10  \bigstrut[t] \\  
P43 &1.50  & 1700  & $8\times 10^{-6}$ & 6   & $1\times 10^{-6}$ & 11  \bigstrut[b]\\
\hline
P44 &1.75  & 1700  & -                 & -    & $7\times 10^{-7}$ & 11  \bigstrut[t]\bigstrut[b]\\
\hline 
P45 &2.00  & 1700  & -                 & -    & $6\times 10^{-7}$ & 11 \bigstrut[t] \bigstrut[b]\\
\hline   
P46 &2.25  & 1700  & -                 & -    & $5\times 10^{-7}$ & 10  \bigstrut[t] \bigstrut[b]\\
\hline
P47 &2.50  & 1700  & -                 & -    & $4\times 10^{-7}$ & 10\bigstrut[t] \bigstrut[b]\\
\hline   
\end{tabular}
\tablefoot{Columns 3--4 lists the dynamical properties for the models with $f_{\mathrm{abs}}=1.0$ and columns 5--6 lists the corresponding properties for the models with $f_{\mathrm{abs}}=0.5$. For the combination of $p$ and $T_{\mathrm{c}}$ where no mass loss or wind velocity is listed, the corresponding model failed to produce an outflow. Note that the resulting wind and dust properties are temporal means as indicated by the angular brackets. }
\end{table}  

This set of dynamical model atmospheres was first presented in Paper I and is here restricted to models that produce outflows (see the colored area in Fig. 4 and 8 of Paper I). All dynamical models in this set have the same stellar parameters: a stellar mass and luminosity of $1\,M_{\odot}$ and $5000\,L_{\odot}$, respectively, an effective temperature of $2800\,$K and solar abundances (corresponding to Model A in set D). The piston velocity amplitude was set to 4 km/s. 

To investigate the dynamical effects of different chemical and optical dust properties we vary the input parameters $p$ and $T_{\mathrm{c}}$ in the formula for the parameterized dust opacity (see Eqs.~\ref{e_fc}-\ref{e_parkap2}). To cover the most probable combinations we let $p$ vary from -1.0 to 2.5 and $T_{\mathrm{c}}$ from $700\,$K up to $1700\,$K, in increments of 0.25 and $100\,$K respectively, resulting in a total of 165 individual dynamical models, of which we select a subsample of 85 models that actually produce outflows. The parameter $\Delta T$ that adjusts the width of the dust formation zone is set to $100\,$K. This value was chosen by comparing with the typical width of the dust formation zone for the models A in set D. In addition to $p$ and $T_{\mathrm{c}}$, we also vary the degree to which the dust opacity is considered true absorption by setting the parameter $f_{\mathrm{abs}}$  to 1.0 and 0.5 respectively (100\% and 50\% true absorption).

The critical dust opacity, $\kcrit=4\pi cGM_\star/L_\star$, needed for the radiative acceleration to overcome the gravitational acceleration is equal to 2.6\,cm$^2$/g for the chosen set of stellar parameters. We set \mbox{$\kappa_{0}=3.0$\,cm$^2$/g} when $p=0$ in Eq.~(\ref{e_pl}) and then adjust for other values of $p$ so that the flux-averaged dust opacity $\kh$ remains fixed. By setting $\kappa_{0}$ such that $\kh>\kcrit$ an outflow will be triggered in these dynamical models if the atmospheric environment is such that it allows for grain growth close enough to the stellar surface ($f_{\mathrm{c}}\approx 1$). 

Table~\ref{t_dynmod} lists the dynamical properties for the individual models in set P. In addition, the middle and bottom panel of Fig. \ref{f_dynobs} show mass-loss rates versus wind velocities for the models with $f_{\mathrm{abs}}=1.0$ and $f_{\mathrm{abs}}=0.5$, respectively. The smaller spread in wind velocity noticeable in the models with a smaller fraction of true absorption ($f_{\mathrm{abs}}=0.5$) is due to a less  pronounced reddening of the radiation field when the dust component absorbs less of the stellar flux (for a detailed discussion on the trends in mass-loss rates and wind velocities for the models in set P, see Paper I).

\section{Spectral synthesis and photometry}
\label{s_modspe}
We calculate spectra and photometry for set D and P, i.e. the dynamical models that include a detailed description of Mg$_2$SiO$_4$ grains (set D) and the models using a parameterized dust description (set P), by sampling 16 snapshots of the radial structure during a typical pulsation cycle, equidistant in phase. The \textit{a posteriori} radiative transfer was computed with opacities from the COMA code \citep{ari00} resulting in an opacity sampling spectrum with a resolution of $\rm R = 10000$, covering the wavelength range between $0.335\,\mu$m and $25\,\mu$m for each snapshot. Based on these data we compute photometric filter magnitudes, following the Bessell system \citep[described in][]{bess88,bess90}, and low resolution spectra ($\rm R = 200$). More details concerning the spectral synthesis can be found in \cite{AGNML09}, where a similar approach was used to obtain the observables of hydrostatic model atmospheres for carbon stars.

In the \textit{a posteriori} radiative transfer the micro-turbulence parameter was set to $\xi = 2.5\,$km/s, which is in agreement with the value used for the gas opacities in the dynamical models. The treatment of the opacities is also consistent with the computation of the atmospheric structures concerning the included atomic, molecular and solid-state sources. For the elemental abundances we assume solar values following \cite{andgrev89}, except for C, N and O where we took the data from \cite{saugrev94}, and we take into account the depletion in the gas phase caused by dust formation. Information about the species included in the radiative transfer, and the line lists, can be found in \cite{AGNML09}. Note that in this work we used only the data of \cite{barb06} for H$_2$O. 

Scattering on dust particles can have a significant effect on the momentum of individual grains and is therefore taken into account when calculating the wind dynamics (see Sect.~\ref{s_detdus}). However, it should be mentioned that we do not consider scattering on dust particles when solving the radiative transfer. For the models in set D the fraction of light scattered by dust particles is small compared to the radiative flux emitted by the star and the scattered photons will not noticeably change the overall radiation field. In fact, scattering of photons in a spherical and unresolved circumstellar shell does not lead to changes in the observed colors, regardless of the  wavelength-depedence of the scattering opacity. This can be understood most easily by considering total energy conservation for a spherical  object observed from arbitrary viewing angles.

\section{Comparative observational data}
\label{s_obsdat}

\begin{figure}
\centering
\includegraphics[width=9cm]{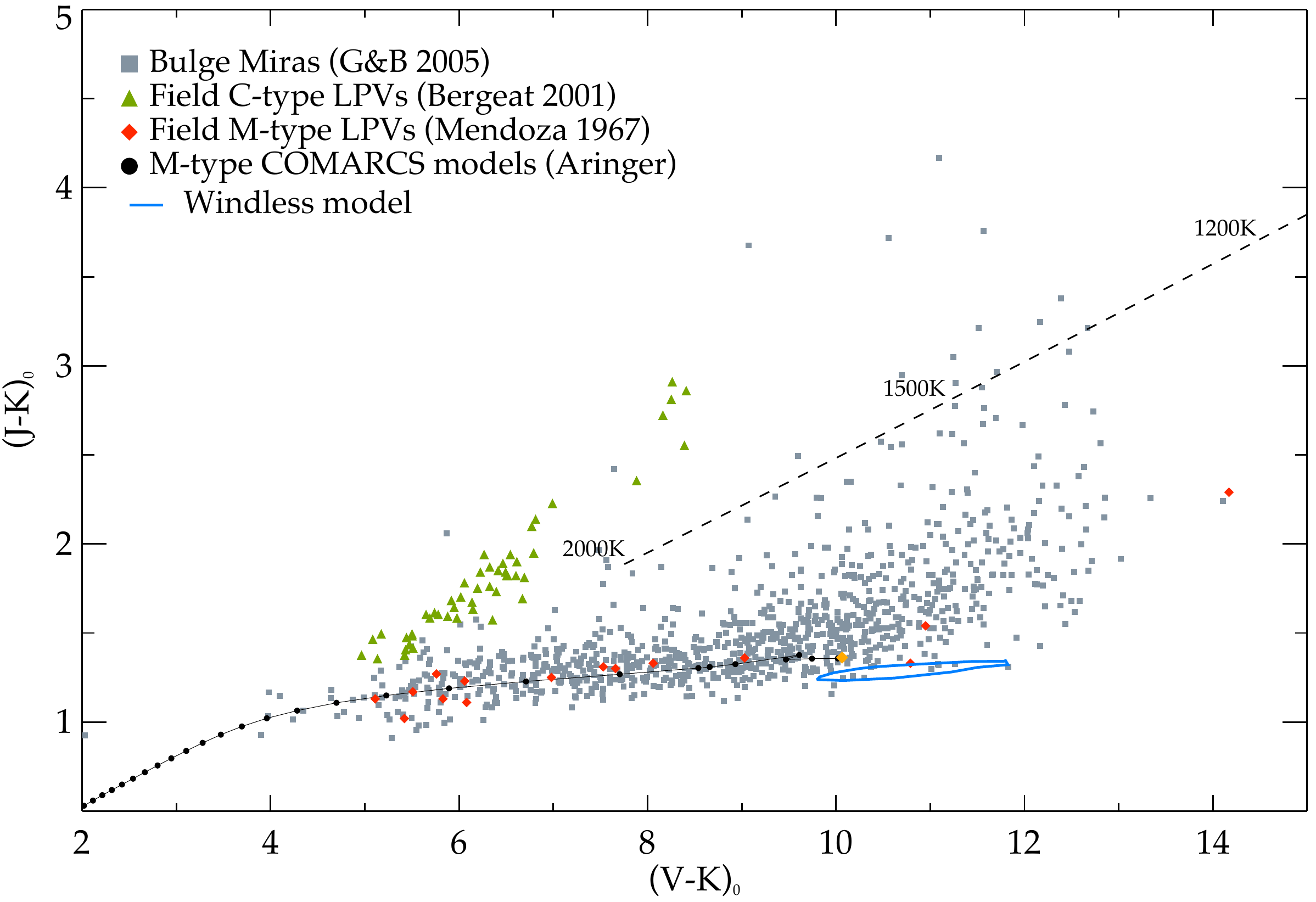}
\caption{Color-color diagram containing observational reference data for comparison with the models, adopted from different sources: Galactic Bulge Miras \citep[grey squares]{gbmiras}, field M-type LPVs \citep[red diamonds]{men67}, and C-rich giants \citep[green triangles]{b01cstars}. Over-plotted are the locations of simulated blackbody emitters in the range of $T_\star=1100-2000$\,K (dashed line), as well as an evolutionary track of a star of solar mass and metallicity, using hydrostatic COMARCS model atmospheres (black circles). The colors of a pulsating atmosphere without wind, with stellar parameters $L_\star=5000$~$L_{\odot}$, $T_\star=2800$\,K and \mbox{$M=1$\,$M_{\odot}$}, are plotted in blue and the static initial model in orange (diamond).}
   \label{f_photogen}
\end{figure}

\begin{figure}
\centering
\includegraphics[width=8.5cm]{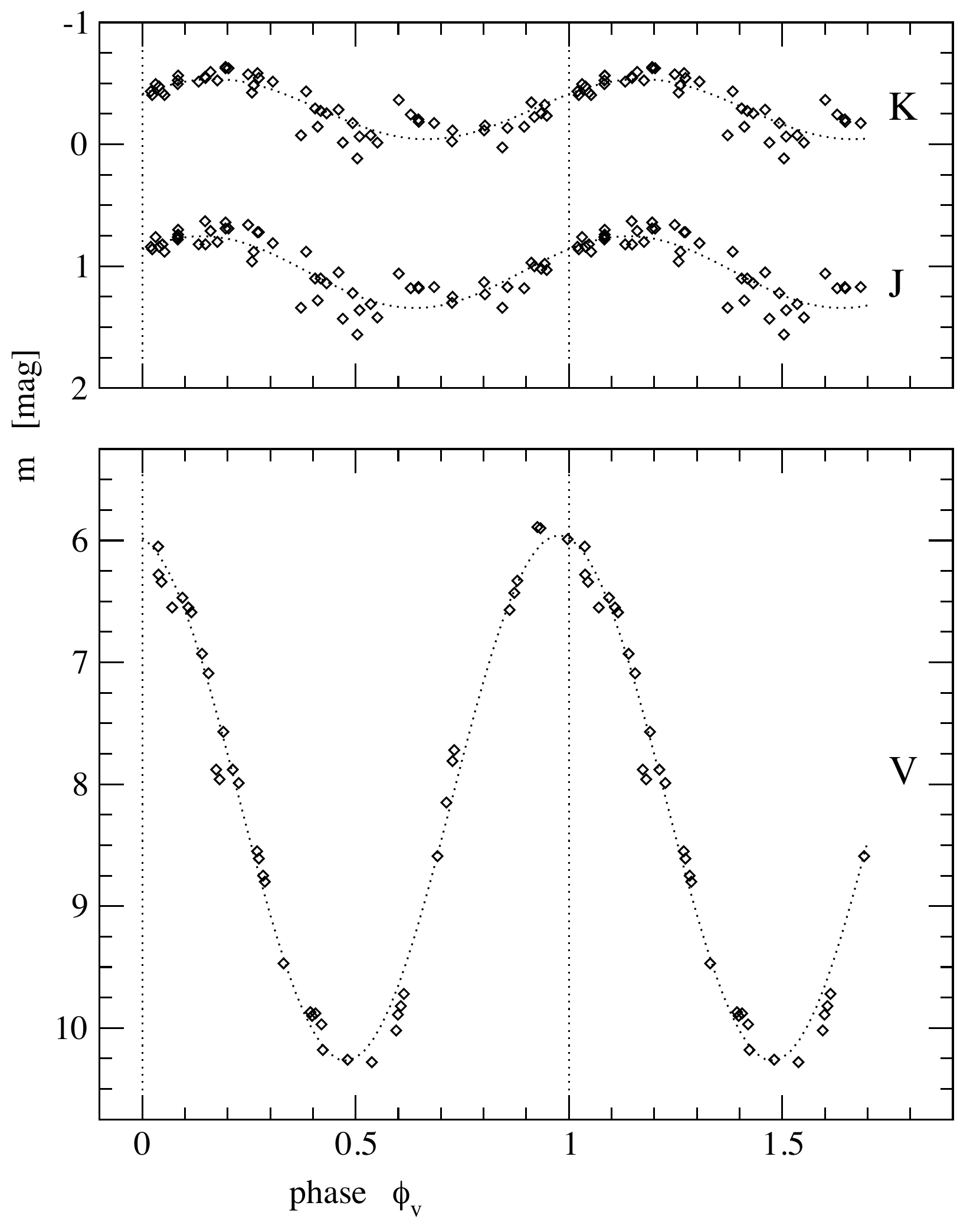}
\caption{Observed photometric variations of the M-type Mira RR\,Sco in different filters. Measurements from different periods are merged into a combined light cycle and each data point is plotted twice to highlight the periodic variations. Sine fits are over-plotted with dotted lines. For references to the observational data, see Sect.~\ref{s_obsdat}.}
   \label{f_lightcurves}
\end{figure}
To gain insight into how different dust properties affect the spectral energy distribution we compare synthetic colors from the sets D and P to observations, not only in the dust-feature rich mid-IR region, but also in the visual and the near-IR part of the spectra. The interest in the visual band is motivated by the observed differences in spectra between \mbox{M-type} and C-type AGB stars in this wavelength region, but also by how different the absorption cross-sections are for the main wind-driving candidates in these two types of stars (i.e. amorphous carbon and Fe-free silicates for C-type and M-type AGB stars, respectively; see Fig.~\ref{f_pl}). Consequently we use the color-color diagram \mbox{($J$\,--\,$K$)} vs. \mbox{($V$\,--\,$K$)} for a detailed analysis of the modeling results and for testing against selected reference observations. Since most AGB stars show noticeable photometric variability, simultaneous observation in all three bands would ideally be required for such a study. However, simultaneous data in the visual and the near-IR are only sparsely available in the literature and we therefore compile data from various sources to estimate \mbox{($V$\,--\,$K$)} values for this color index. Figure~\ref{f_photogen} shows the resulting compilation. 

The largest and most important group of targets in this figure consists of M-type Miras identified in the galactic bulge by \citet{gbmiras} while exploiting the photometric variations provided by the OGLE survey. The authors cross-correlated their list of Miras with the 2MASS catalogue and provide the resulting single-epoch $JHK$ photometry. The latter is used in this work after de-reddening according to \citet{MatFN05} and converting the magnitudes to the Bessell system following \citet{Carpe01}. In addition, Groenewegen \& Blommaert (priv. comm.) cross-correlated their target list with the OGLE ``photometric maps'' \citep{USKPS02} and provided mean $V$ magnitudes, which we de-reddened according to \citet{Sumi04}. Both kinds of data can subsequently be merged to estimate \mbox{($V$\,--\,$K$)} colors. Note that the flux variations in the $K$ band are much smaller than in the $V$ band (see, e.g., Fig.~\ref{f_lightcurves}) and the uncertainties of color indices based on single-epoch data for the $K$ magnitudes should be within limits.

A few other data sets are also shown in Fig.~\ref{f_photogen}. \citet{men67} presented multi-color photometry for a number of LPVs and we adopted values for the M-type subsample from their Table\,3 and transformed the magnitudes following the relations given in \citet{bess88}. For illustration we also over-plotted the observed photometry compiled by \citet{b01cstars}, representing C-rich giants with moderate properties concerning variability and mass loss \citep[cf.][]{now11}. Additionally, we show the evolutionary track of a typical solar-like star \mbox{(1\,$M_{\odot}$, $Z_{\odot}$)} based on synthetic evolutionary models \citep{Marigo07,Marigo08} and hydrostatic COMARCS model atmospheres \citep{AGNML09}, using the same approach to compute synthetic photometry as in this work. 

Most evolved AGB stars belong to the group of LPVs and the photometric magnitudes and colors -- e.g., the ones in Fig.~\ref{f_photogen} -- change with pulsation phase. A comparison of the light variations of observed targets with the corresponding modeling results would be very valuable for constraining the modeling method. Unfortunately, no systematic photometric monitoring of M-type LPVs simultaneously covering visual and near-IR magnitudes has been performed so far. We make an attempt to remedy this situation by using the same approach as outlined in detail in \citet{now11} for the C-type Mira RU\,Vir. In order to use this method the targets must have reasonably regular light-curves (which we checked with the help of data from the AAVSO database\footnote{http://www.aavso.org/data/lcg}) and time-series photometry in various bands should be available. We ended up with a list of seven M-type Miras (R\,Car, R\,Hya, R\,Oct, R\,Vir, RR\,Sco, T\,Col, and T\,Hor). For these objects we adopted photometric data in the visual from \citet{Eggen75a} as well as \citet{men67}, and complemented those with the near-IR data published by \citet{WhiMF00}. The observed photometric measurements were combined into one composite light cycle as described in \citet{now11}, the phase information needed for this merging was again determined on the basis of AAVSO visual light-curves. This resulted in combined light-curves as can be seen in Fig.~\ref{f_lightcurves}. Based on a sine fit to the latter we can derive characteristic photometric variations, even for filter combinations where no simultaneous observations were obtained, as for example the desired \mbox{($V$\,--\,$K$)} color. The loops produced by the selected M-type Miras in the color-color diagram chosen for our comparison can be seen in the top panels of Fig.~\ref{f_photoobs}. To illustrate the strong difference in the visual and near-IR colors, we also include the results for the C-type Mira RU\,Vir from \citet{now11}.

\section{Photometry of the models in set D}
\label{s_phdetdus}

\begin{figure*}
\centering
\includegraphics[width=17cm]{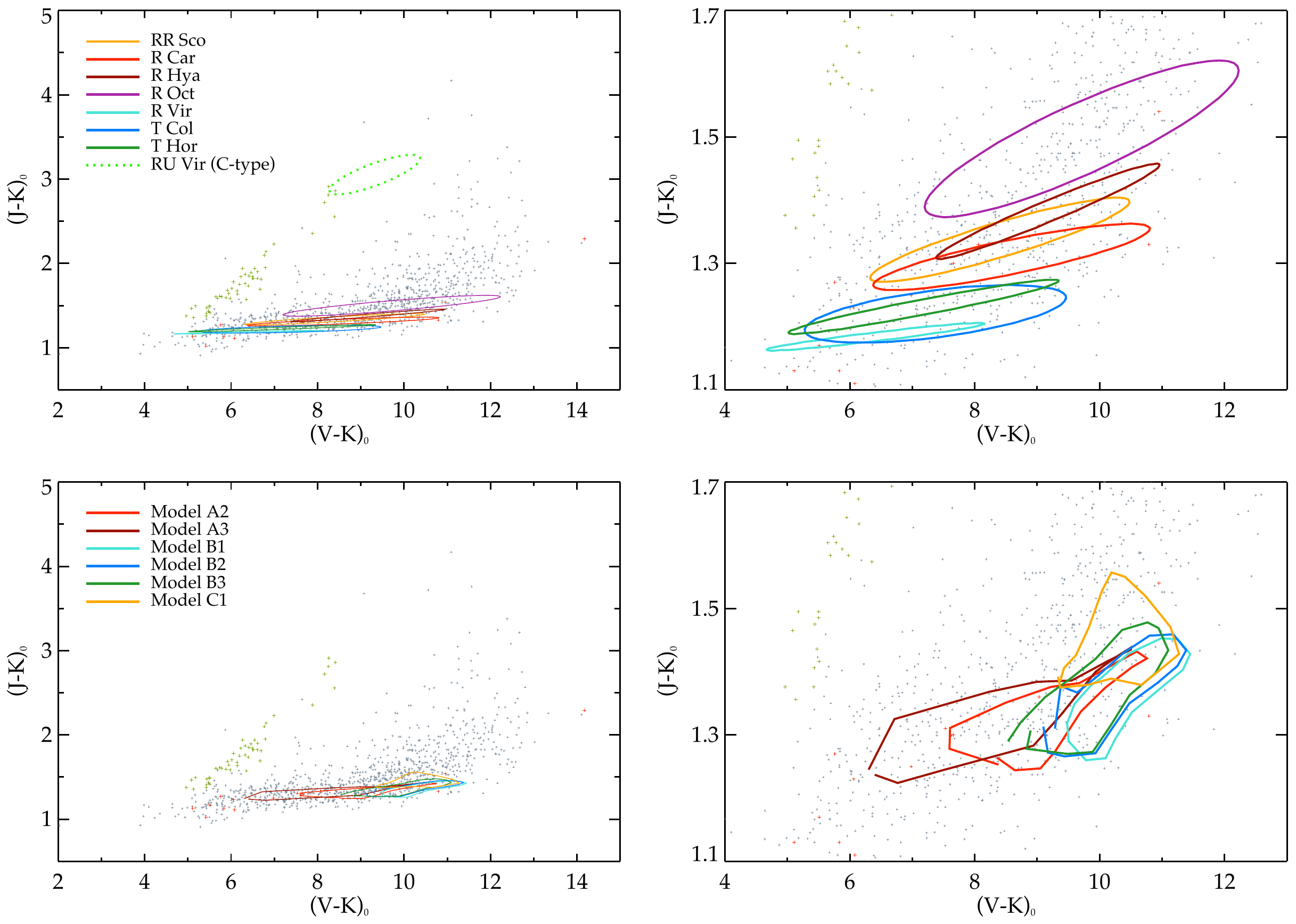}
\includegraphics[width=17cm]{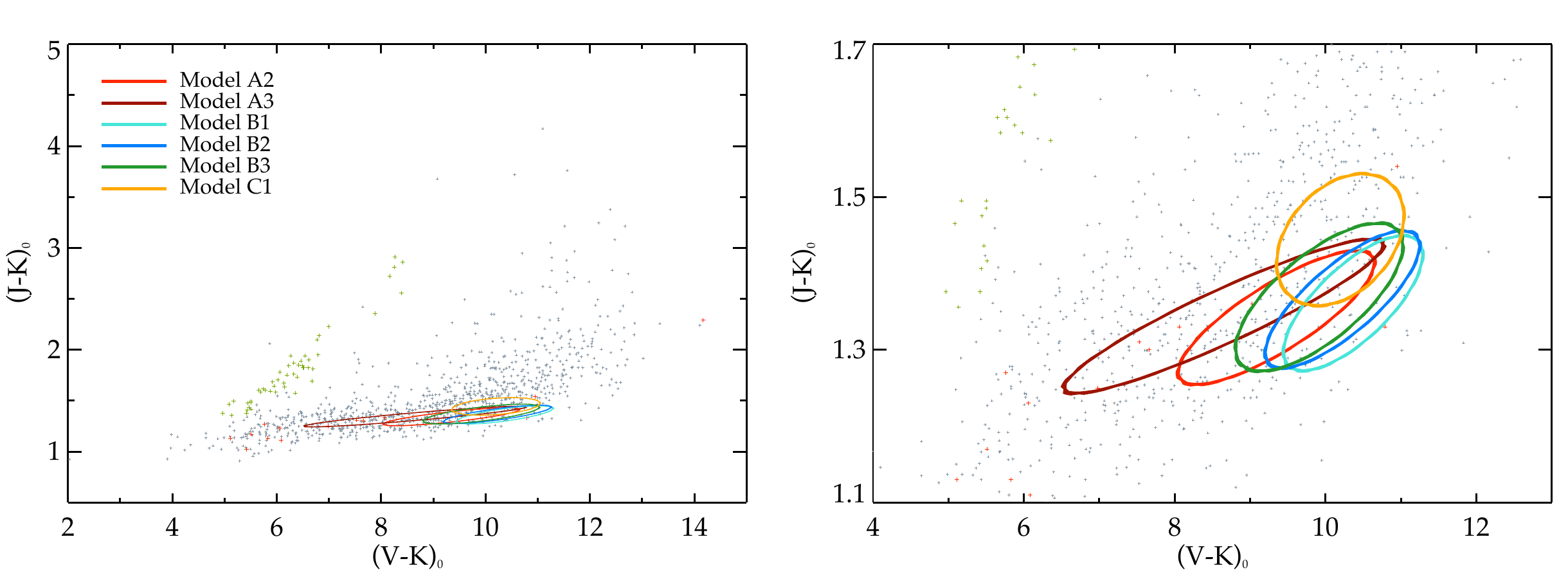}
\caption{Observed and synthetic photometric variations of M-type Miras (and one C-type Mira RU\,Vir, which is included for illustration purposes). \textit{Top panels:} Photometric variations for the sample of observed targets, derived from sine fits of light-curves (see text in Sect.\,\ref{s_obsdat} for details). \textit{Middle panels:} Photometric variations for the dynamical models in set D. \textit{Bottom panels:} Photometric variations for the dynamical models in set D, with colors calculated from sine fits of the light-curves, same as for the observational data in the top panels. The right panels show the same content, zoomed in and centered on the color loops. For comparison, we also show the observational data of Fig.~\ref{f_photogen}}
\label{f_photoobs}
\end{figure*}

\begin{figure*}
\centering
\includegraphics[width=8.5cm]{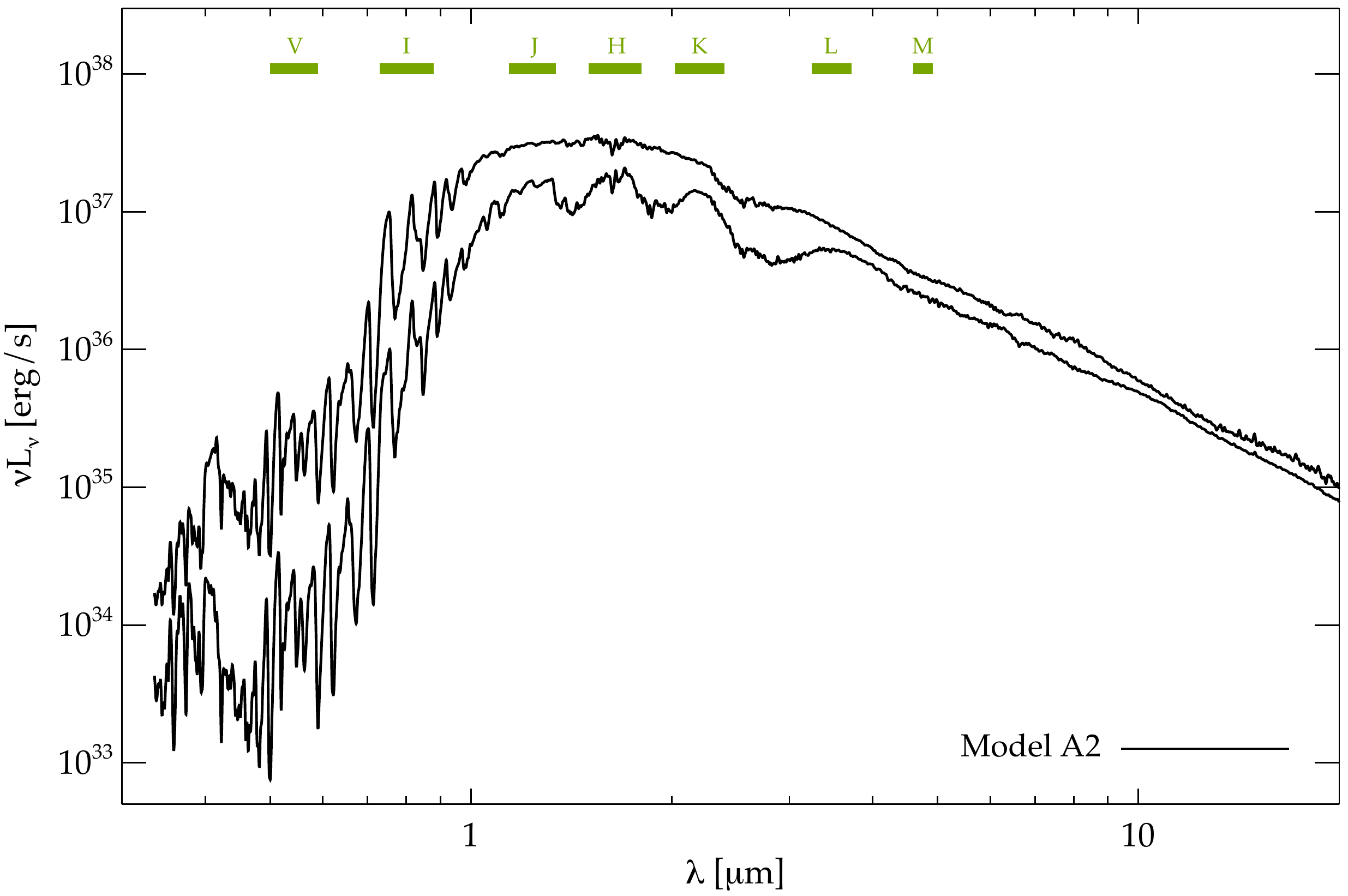}
\includegraphics[width=8.5cm]{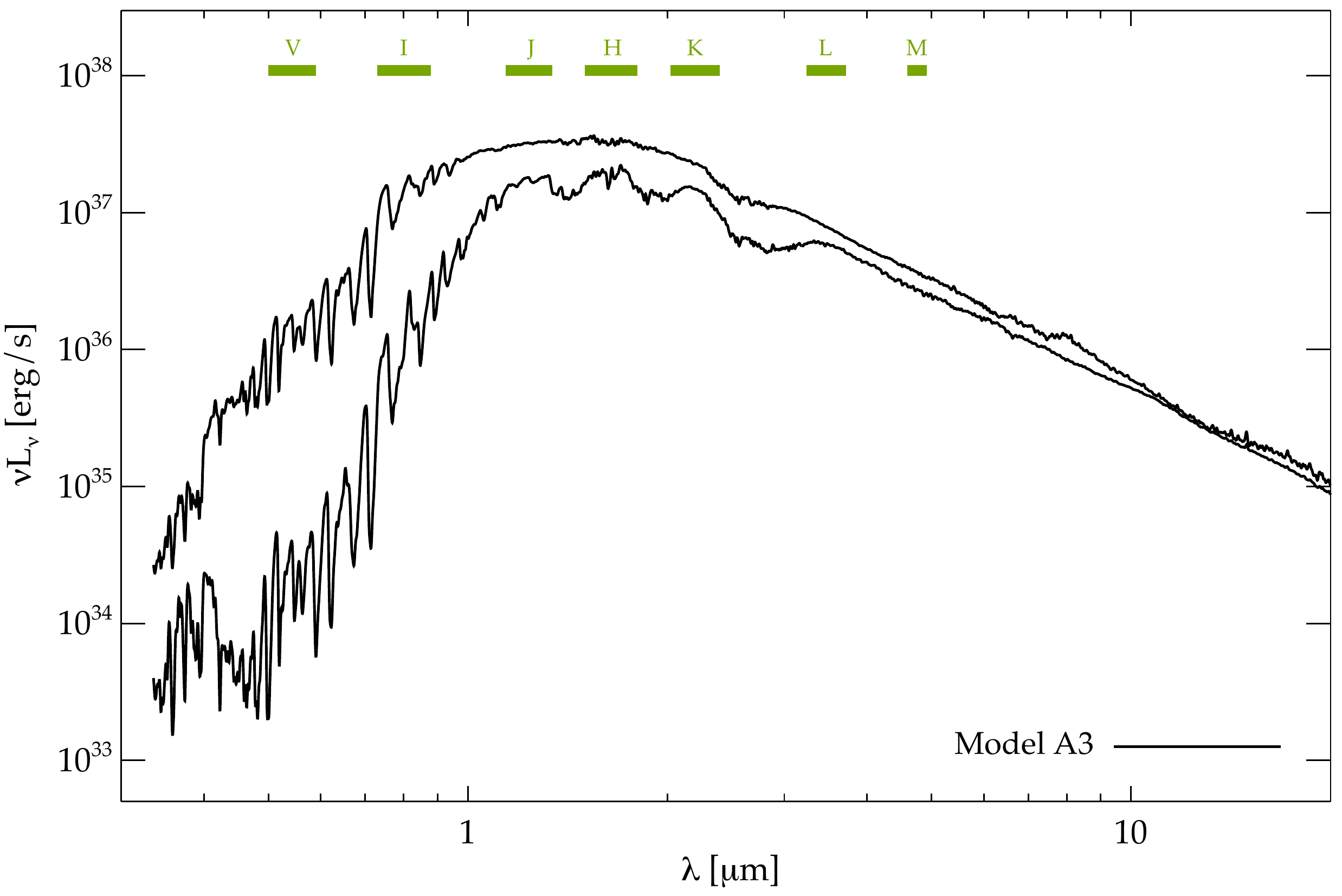}
\includegraphics[width=8.5cm]{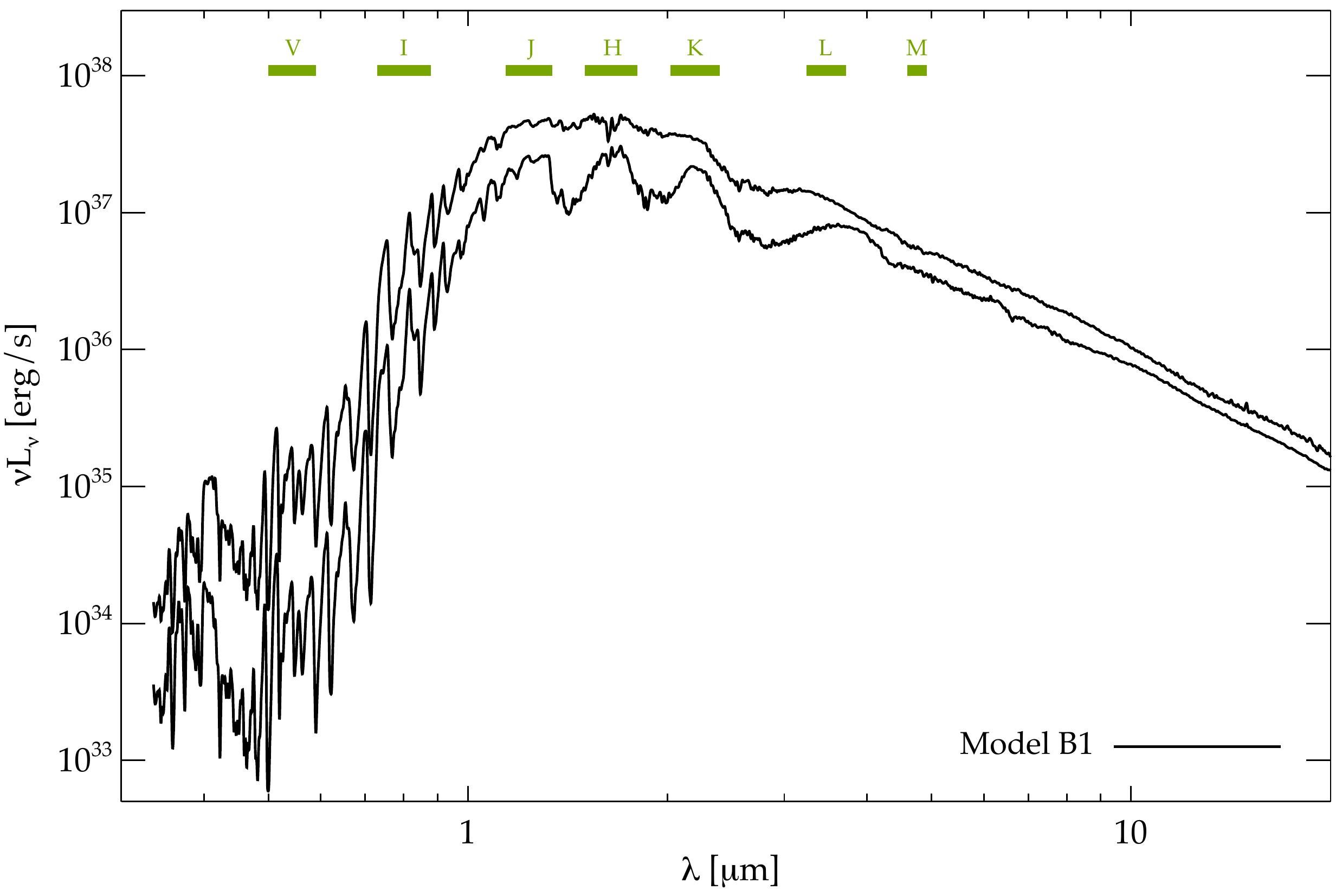}
\includegraphics[width=8.5cm]{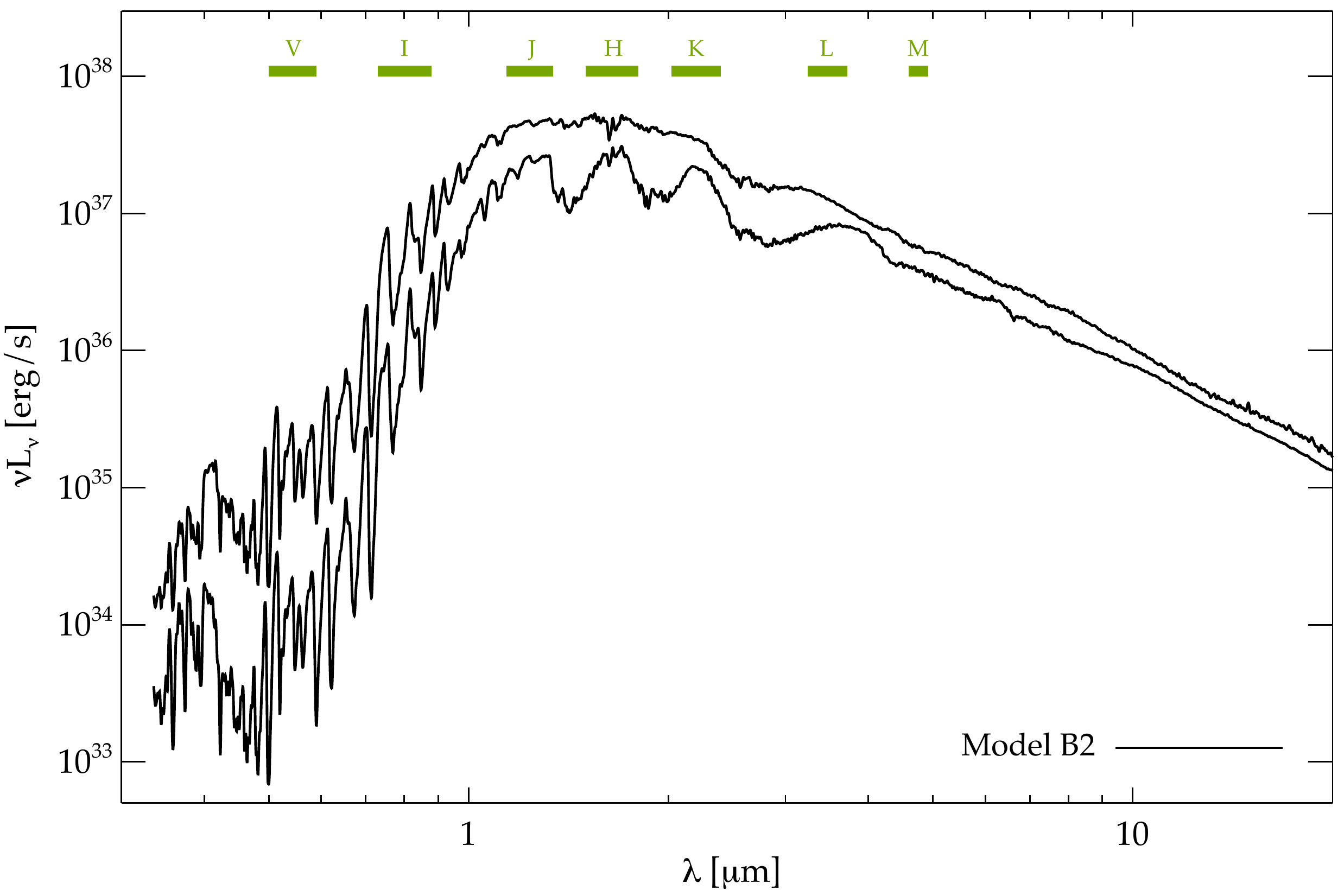}
\includegraphics[width=8.5cm]{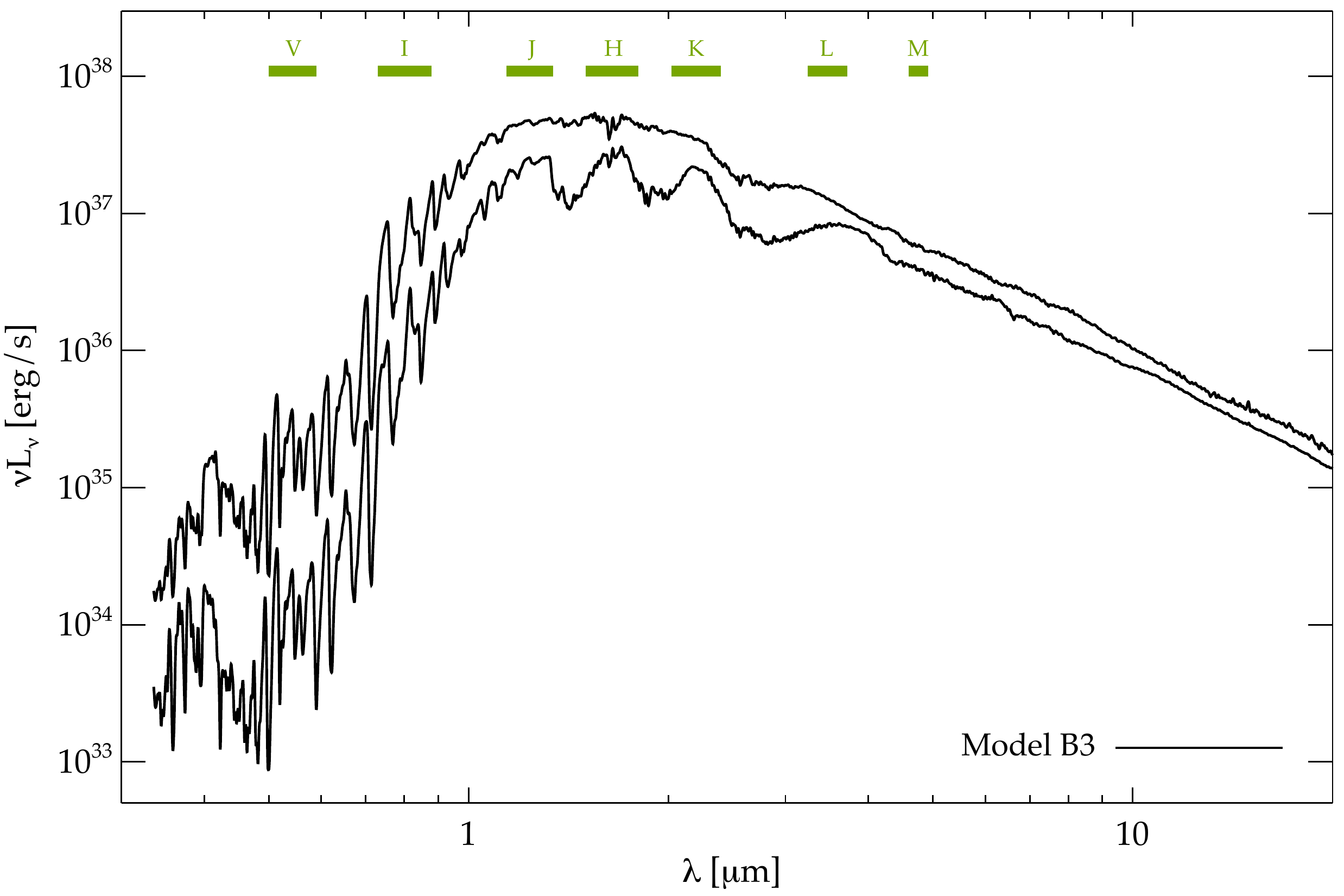}
\includegraphics[width=8.5cm]{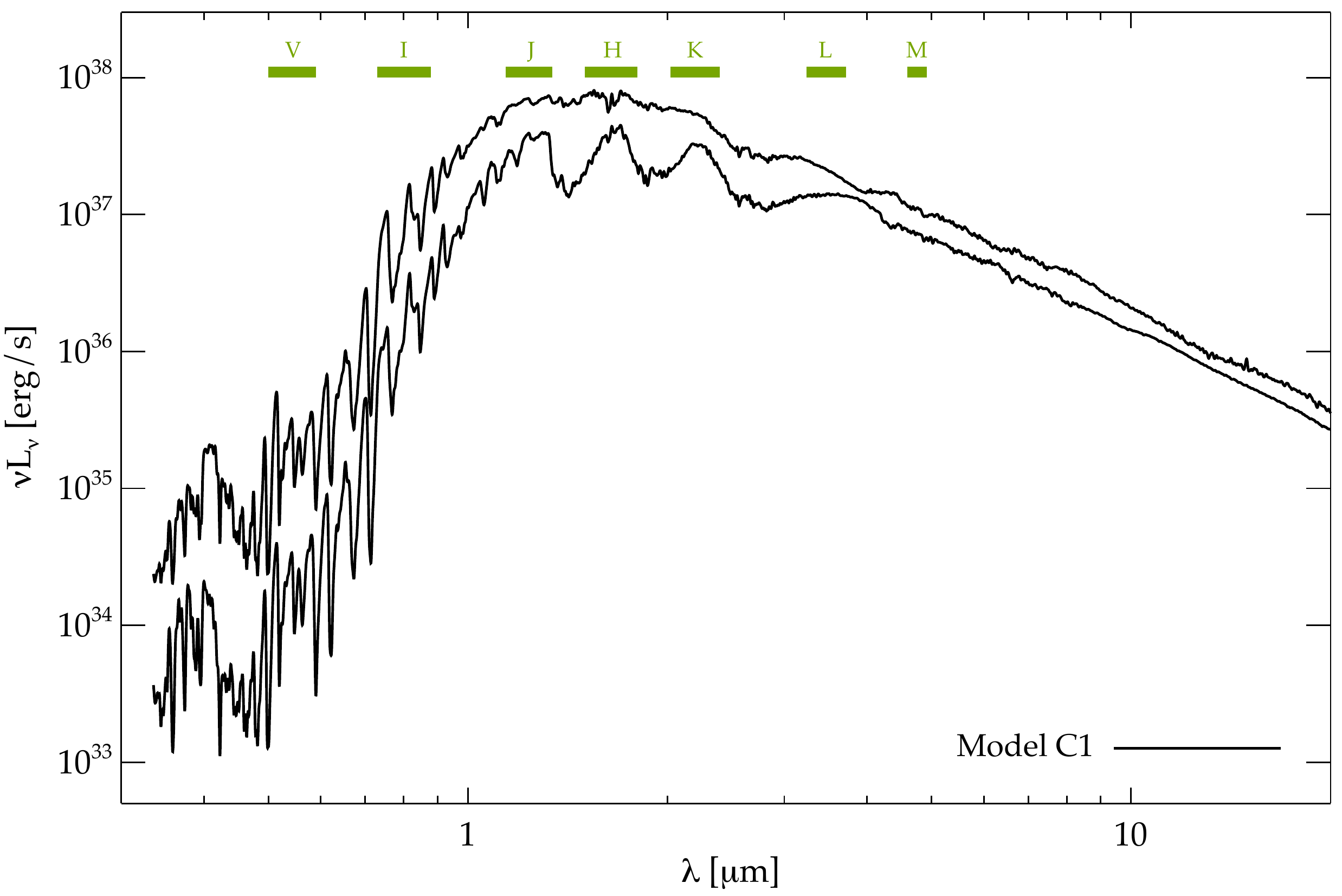}
   \caption{ Spectra of the DMAs in set D, during a luminosity maximum (upper curve) and a luminosity minimum (lower curve). For identification of molecular features, see Fig.~\ref{f_species}.}
    \label{f_spectra}
\end{figure*}

Realistic dynamical model atmospheres of M-type AGB stars should be able to capture the time-dependent photometric variations seen in the top panels of Fig.~\ref{f_photoobs}, something hydrostatic models are unable to do, even if they might produce similar colors as a variable object at a given time-instance. In order to compare photometry from our dynamical models with observed photometric variations we calculate synthetic spectra, and subsequently photometric fluxes, for the wind models in set D, covering a full pulsation cycle. The resulting colors can be seen in the middle panels of Fig.~\ref{f_photoobs}.  The lower panels of Fig.~\ref{f_photoobs} also show the colors for the models in set D, but in this case calculated from sine fits of the synthetic light-curves in the $V$, $J$ and $K$ bands. This is done in the same way as for the observational data (see Sect.~\ref{s_obsdat} and Fig.~\ref{f_lightcurves}) in order to facilitate the comparison between the observed and synthetic photometry. 

As demonstrated by Fig.~\ref{f_photoobs}, the models in set D show similar behavior as the observed targets in the colors ($J$\,--\,$K$) and ($V$\,--\,$K$), with loops occupying effectively the same range in the color-color diagram. Most noticeable, both observed and synthetic photometry show large variations in \mbox{($V$\,--\,$K$)} and small variations in \mbox{($J$\,--\,$K$)} during a pulsation cycle. A closer examination of the synthetic photometry reveals that the variations in \mbox{($V$\,--\,$K$)} are largest for the dynamical models with stellar parameters according to combination A, followed by B and then C. The variations in ($J$\,--\,$K$) are very similar for all models, except for the model C1. This is a consequence of the high mass-loss rate combined with the relatively low wind velocity, resulting in a higher density in the wind-acceleration zone than for the models A and B (since $\dot{M}\propto\rho u$). A higher density will cause more of the stellar radiation to be thermally reprocessed by the circumstellar matter, and consequently, slightly redder values for \mbox{($J$\,--\,$K$)}.

To understand what is causing the temporal variations in color, we plot the spectra of the models in set D during luminosity extremes (minimum and maximum, respectively) in Fig.~\ref{f_spectra}. It is clear from these spectra that the flux variations during luminosity extremes are larger in the $V$ band than in the $J$ and $K$ band and that the variations in \mbox{($V$\,--\,$K$)} will be dominated by the changes in the visual flux. The flux variations in the $J$ and $K$ bands during luminosity extremes are of the same magnitude, leading to small variations in \mbox{($J$\,--\,$K$)}. 

In the visual wavelength region the molecular features are mostly due to TiO, whereas H$_2$O features have a strong impact in the near-IR wavelength region (see the relevant molecular contributions in Fig.~\ref{f_species}). However, H$_2$O does not produce any deep features in the $K$ and $J$ band since the filters are designed to avoid atmospheric water vapor. The strong influence of TiO on the $V$ band, on the other hand, is noticeable in the spectra shown in Fig~\ref{f_spectra}. The variations in the visual band are most pronounced in the models A2 and A3. Figure~\ref{f_press} shows the abundance of TiO as a function of distance from the star, indicating the distance where the optical depth $\tau=2/3$ in the $V$ band with vertical lines.\footnote{The distance from the star where $\tau=2/3$ for the visual band is estimated by \mbox{$R_V(\tau=2/3)=\int F_V(\lambda)R_{\lambda}(\tau=2/3)d\lambda/\int F_V(\lambda)d\lambda$}. Here $F_V$ is the filter function for the $V$ band and $R_{\lambda}$ is the distance from the star where $\tau=2/3$ for a wavelength $\lambda$.} It is clear that the abundance of TiO} varies strongly at this distance during a pulsation cycle in model A3, while no significant variation occurs in model B3 and C1. The lower concentration of TiO during the luminosity maximum, and consequently the larger variations in \mbox{($V$\,--\,$K$)}, is caused by the higher effective temperature of model A3 (see Table~\ref{t_parmodbg}).

As mentioned in Sect.~\ref{s_mdetdust}, the outflows of the models in set D are driven by photon scattering on Mg$_2$SiO$_4$ grains with a very low absorption cross-section in the visual and near-IR. The scattering is predominantly forward oriented for the grain sizes produced in these models (see Fig.~\ref{f_gsca}), and the spectral energy distribution will not be affected significantly. As a result we see strong molecular features in the visual band and weak circumstellar reddening, i.e. not much of the stellar radiation is thermally reprocessed by the dust particles. Consequently, the transparent circumstellar envelope will lead to small variations in \mbox{($J$\,--\,$K$)} and large variations in \mbox{($V$\,--\,$K$)}. The striking similarity between the observed photometric variations and the synthetic photometry gives support for stellar winds driven by photon scattering on dust grains. The fact that hydrostatic COMARCS models (see Fig~\ref{f_photogen}), as well as dynamical model atmospheres without wind \citep[e.g.][]{tej03}, produce colors very similar to observed M-type AGB stars may be taken as another indication that the circumstellar envelopes are quite transparent. This is not the case for C-type AGB stars, where objects with even moderate mass-loss rate show noticeable circumstellar reddening due to dust absorption \citep[e.g.][]{now11}.

\begin{figure*}
\centering
\includegraphics[height=5.5cm]{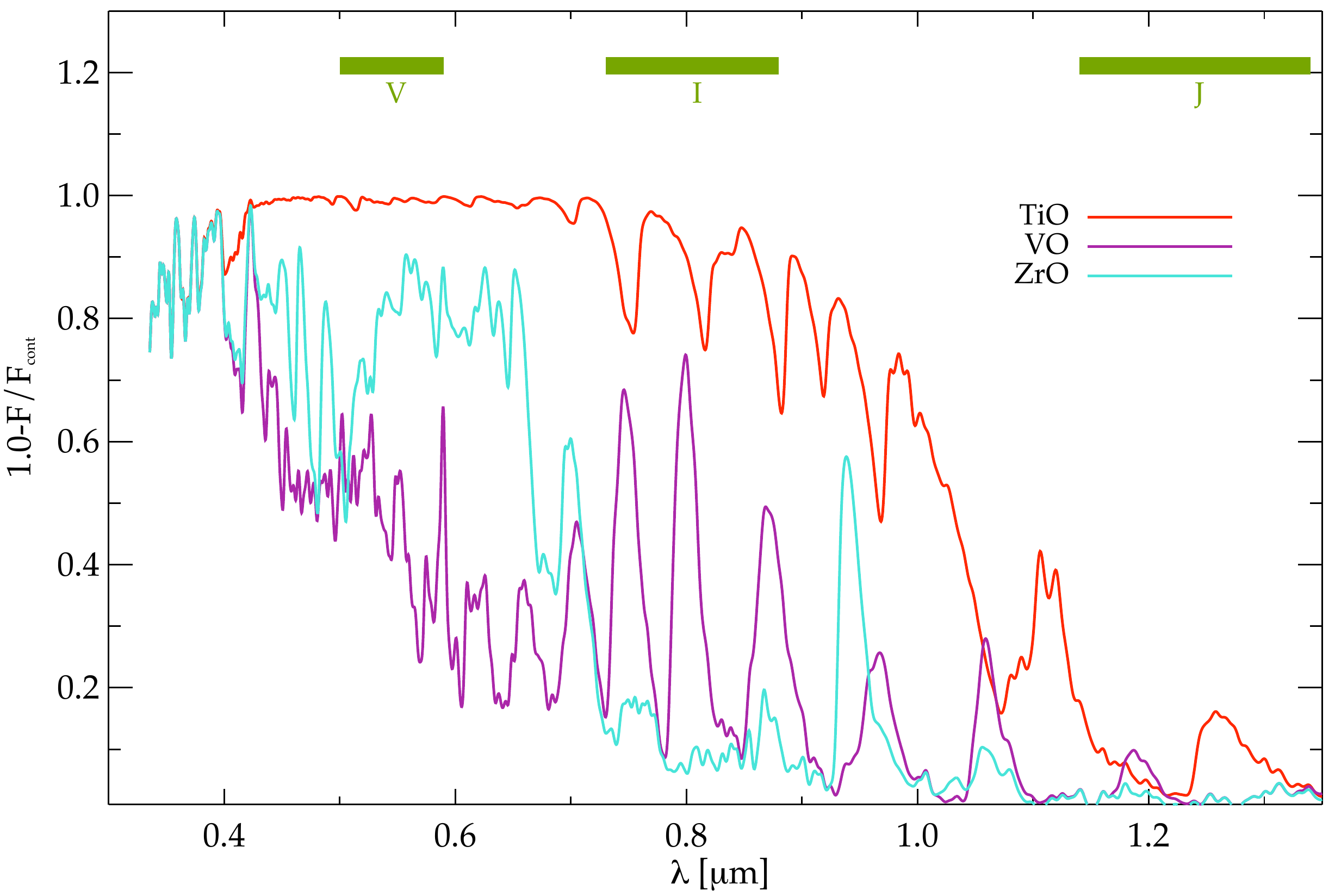}
\includegraphics[height=5.5cm]{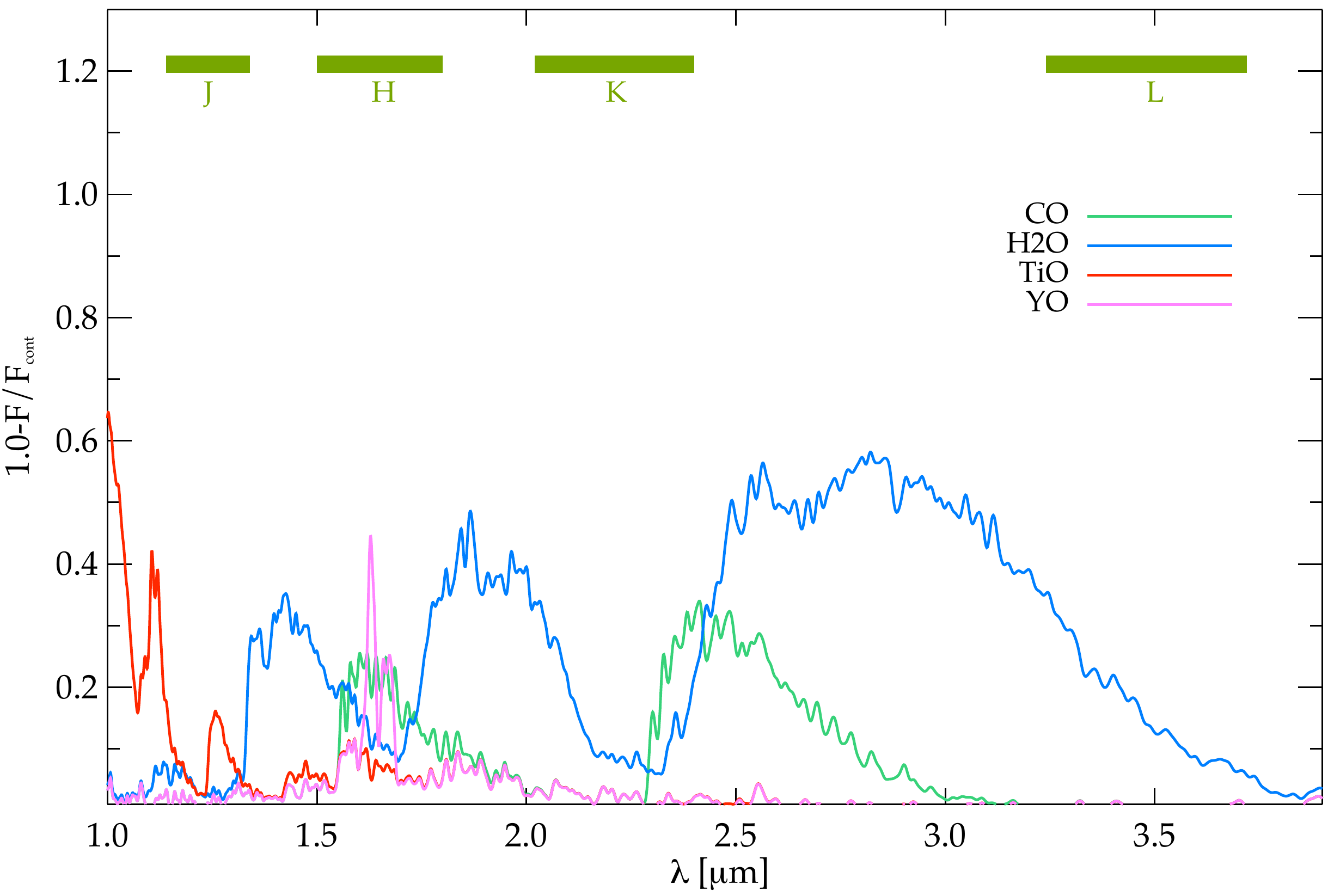}
   \caption{Important molecular contributions in the spectrum of an M-type AGB star, in the wavelength region 0.3-1.35 $\mu$m (left panel) and 1.0-3.9 $\mu$m (right panel), calculated for a hydrostatic model atmosphere with $L=5000$~$L_{\odot}$, $T_\star=2800$\,K, \mbox{$M=1$\,$M_{\odot}$.} The different photometric bands are also marked.}
    \label{f_species}
\end{figure*}

\begin{figure}
\centering
\includegraphics[width=8.5cm]{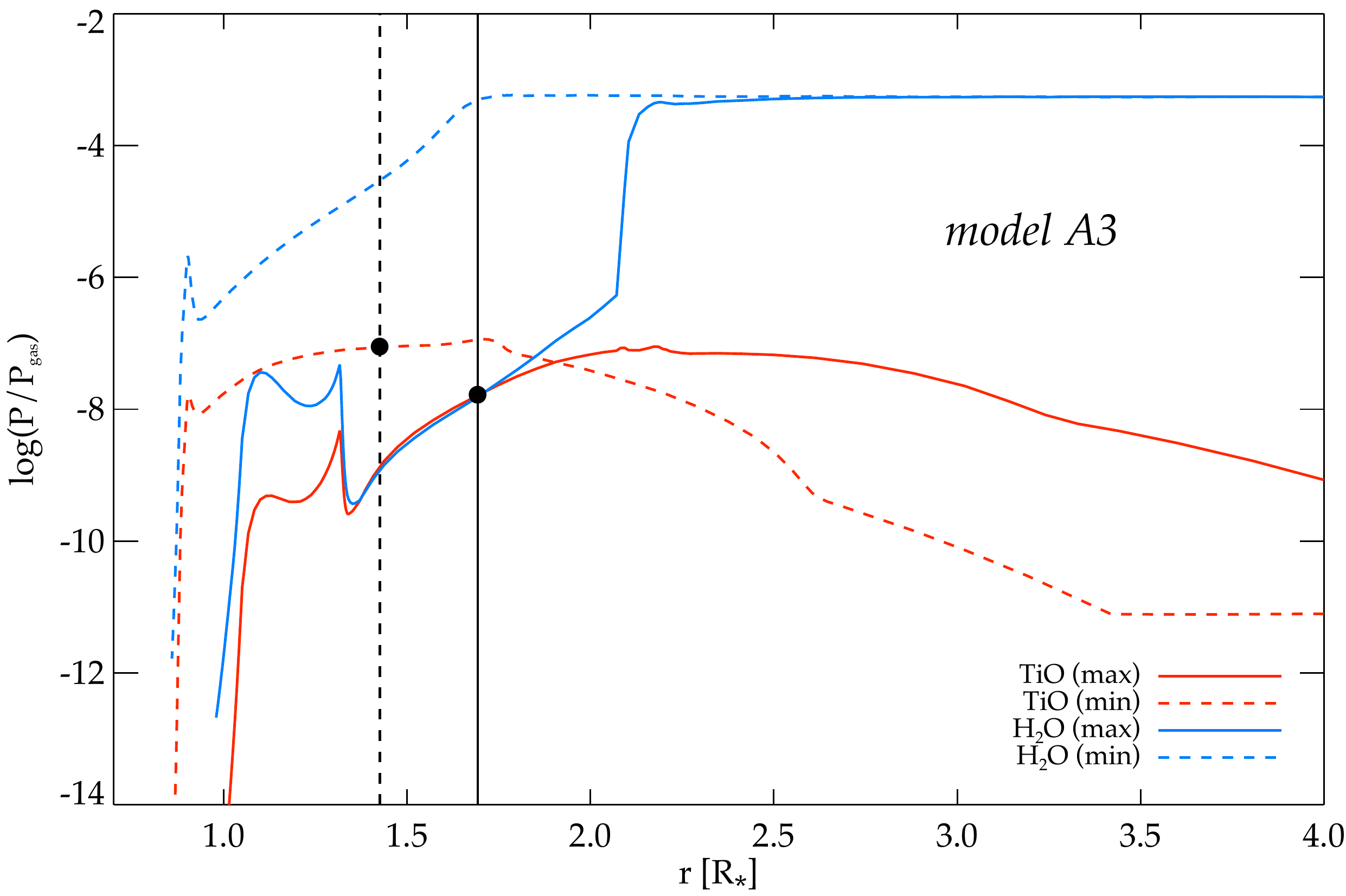}
\includegraphics[width=8.5cm]{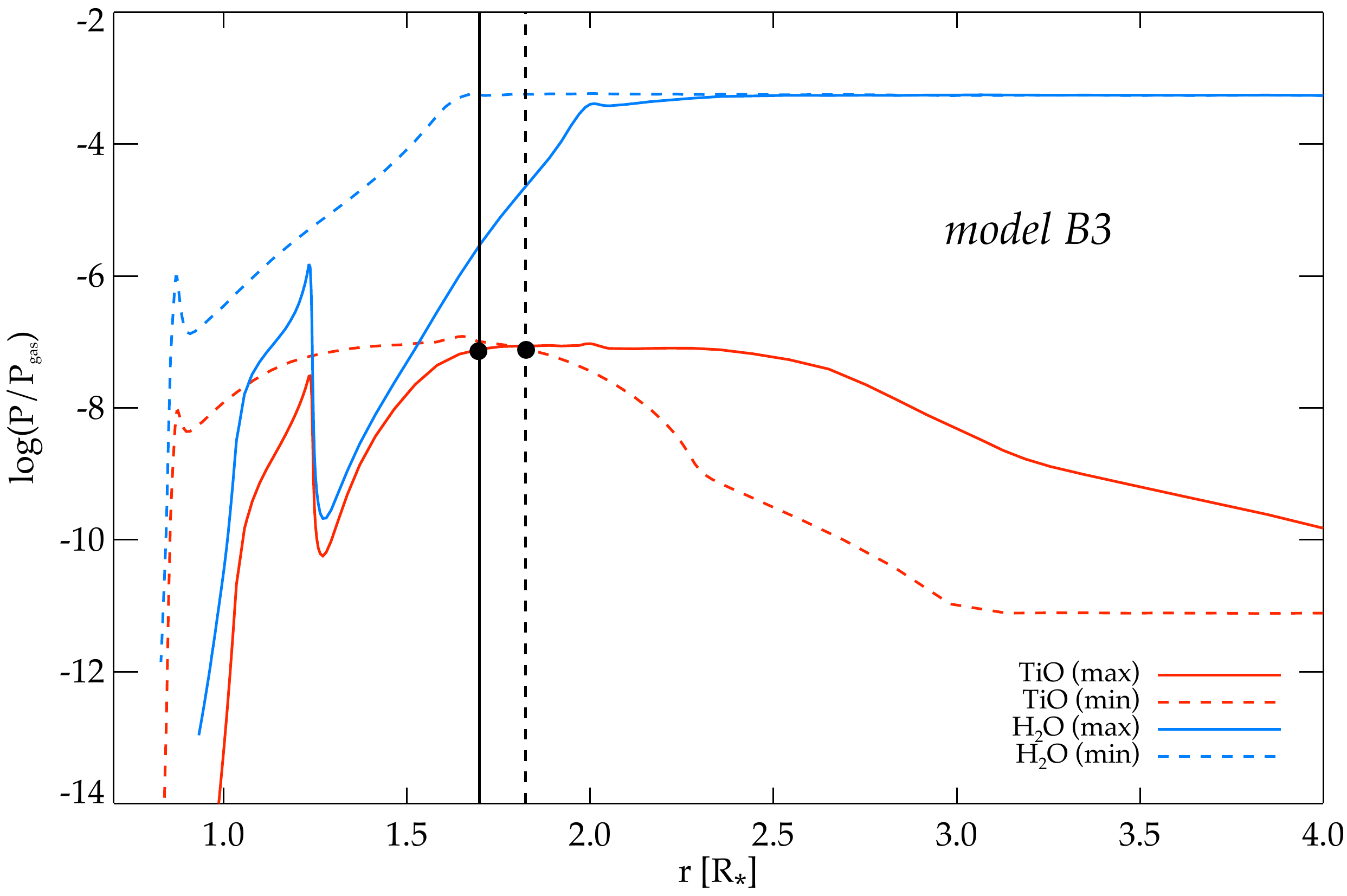}
\includegraphics[width=8.5cm]{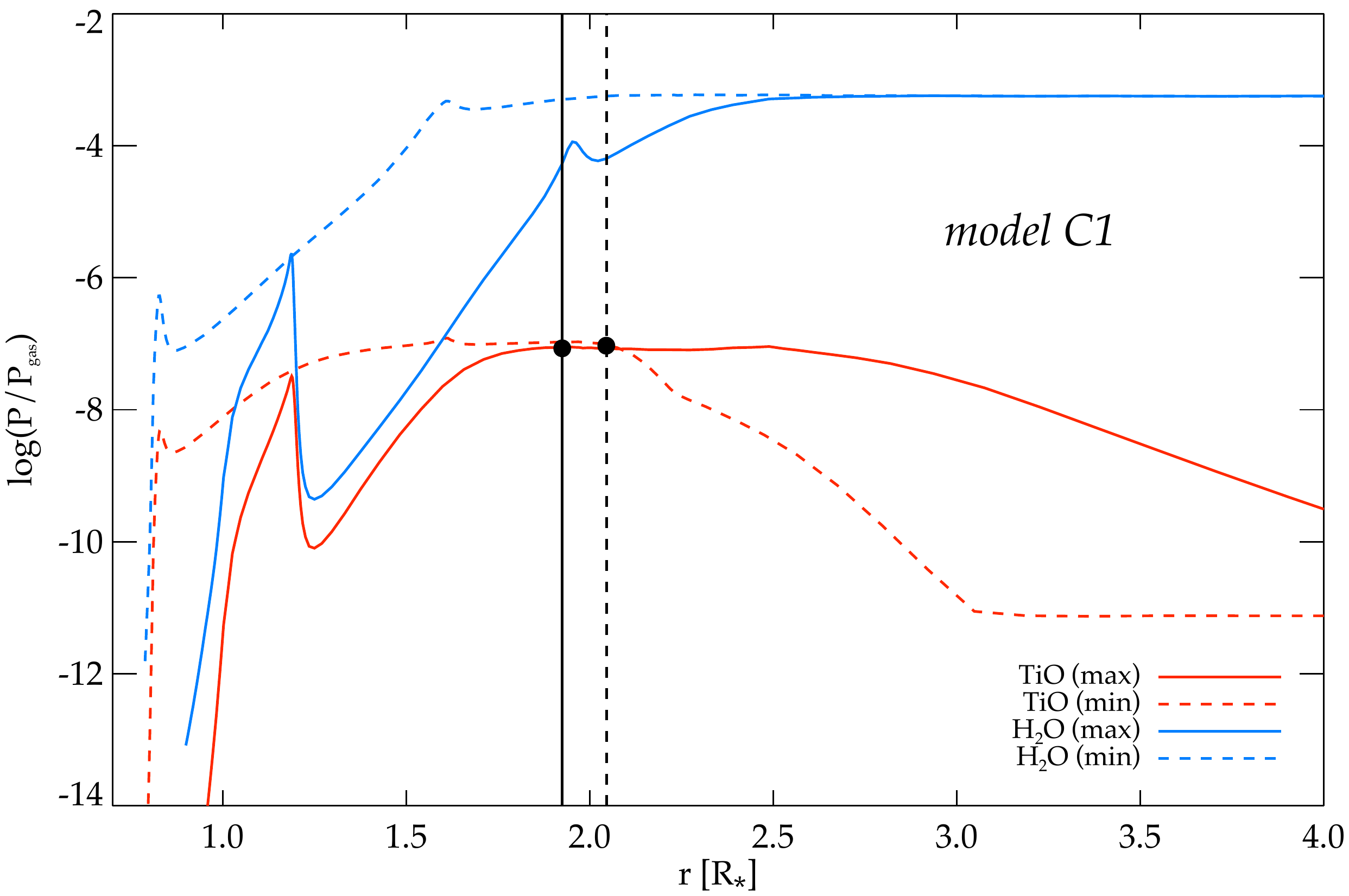}
   \caption{Abundance of TiO (red) and H$_2$O (blue) as a function of distance from the star for the models A3, B3 and C1, during minimum (dashed lines) and maximum (solid lines) luminosity. The vertical lines mark the average distance where $\tau=2/3$ for the $V$ band during minimum (dashed line) and maximum (solid line) luminosity. Since TiO features dominate in the visual wavelength region the black circles mark the abundance of interest for the $V$ band (note that H$_2$O does not contribute to this band).}
    \label{f_press}
\end{figure}

\section{Photometry of the models in set P}
\label{s_phpardus}
At this point the question arises if the shape, location and tilt of the photometric variations discussed above are generic properties of dynamical models, or to what degree they do give constraints on the chemical composition and absorption cross-section of the grain material driving the wind. In order to investigate how photometric data is influenced by different optical and chemical properties of a hypothetical wind-driving dust species, we calculate photometry and spectra for the dynamical models in \mbox{set P.}

\subsection{Phase-averaged photometry}
\begin{figure}
\centering
\includegraphics[width=8.5cm]{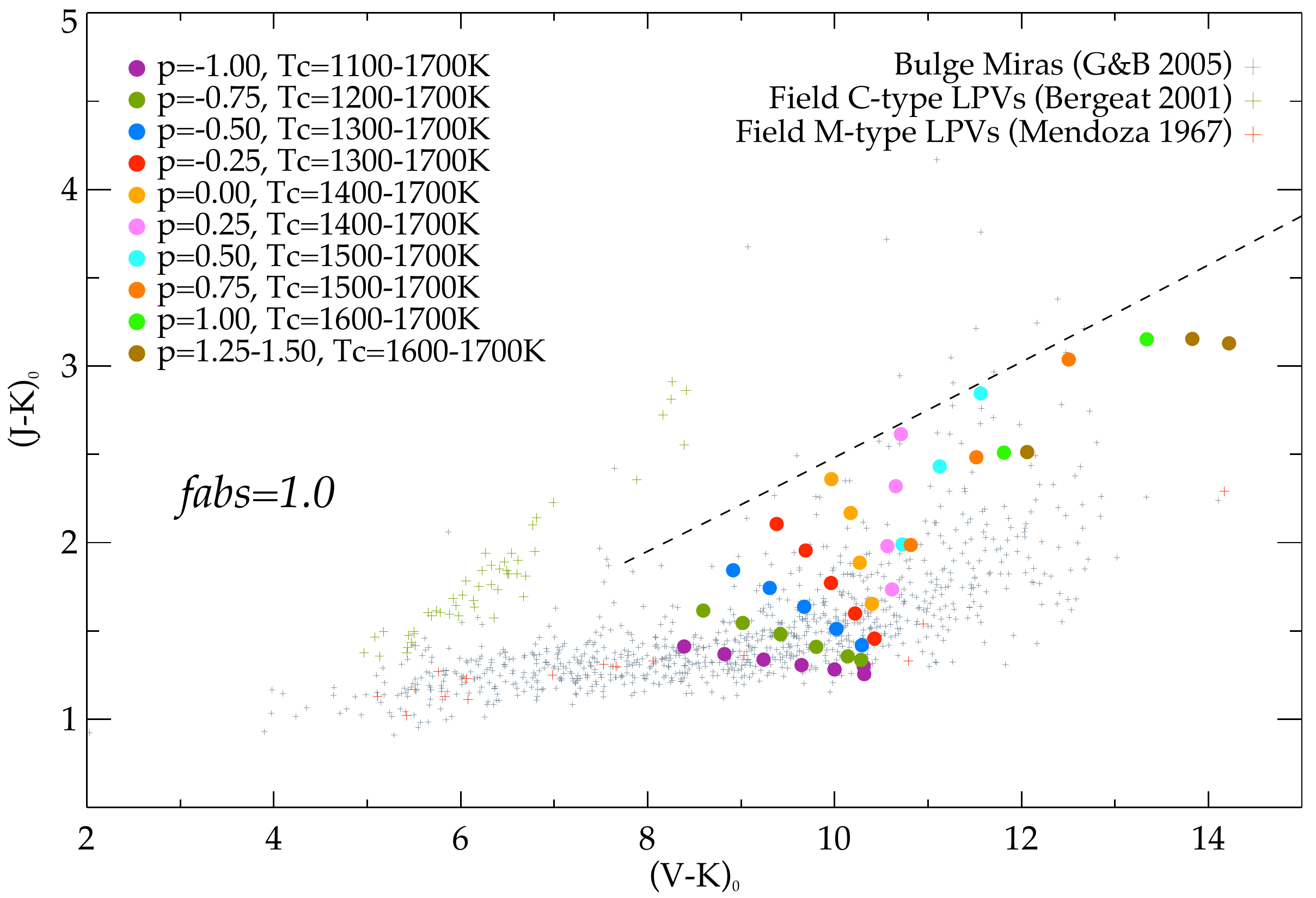}
\includegraphics[width=8.5cm]{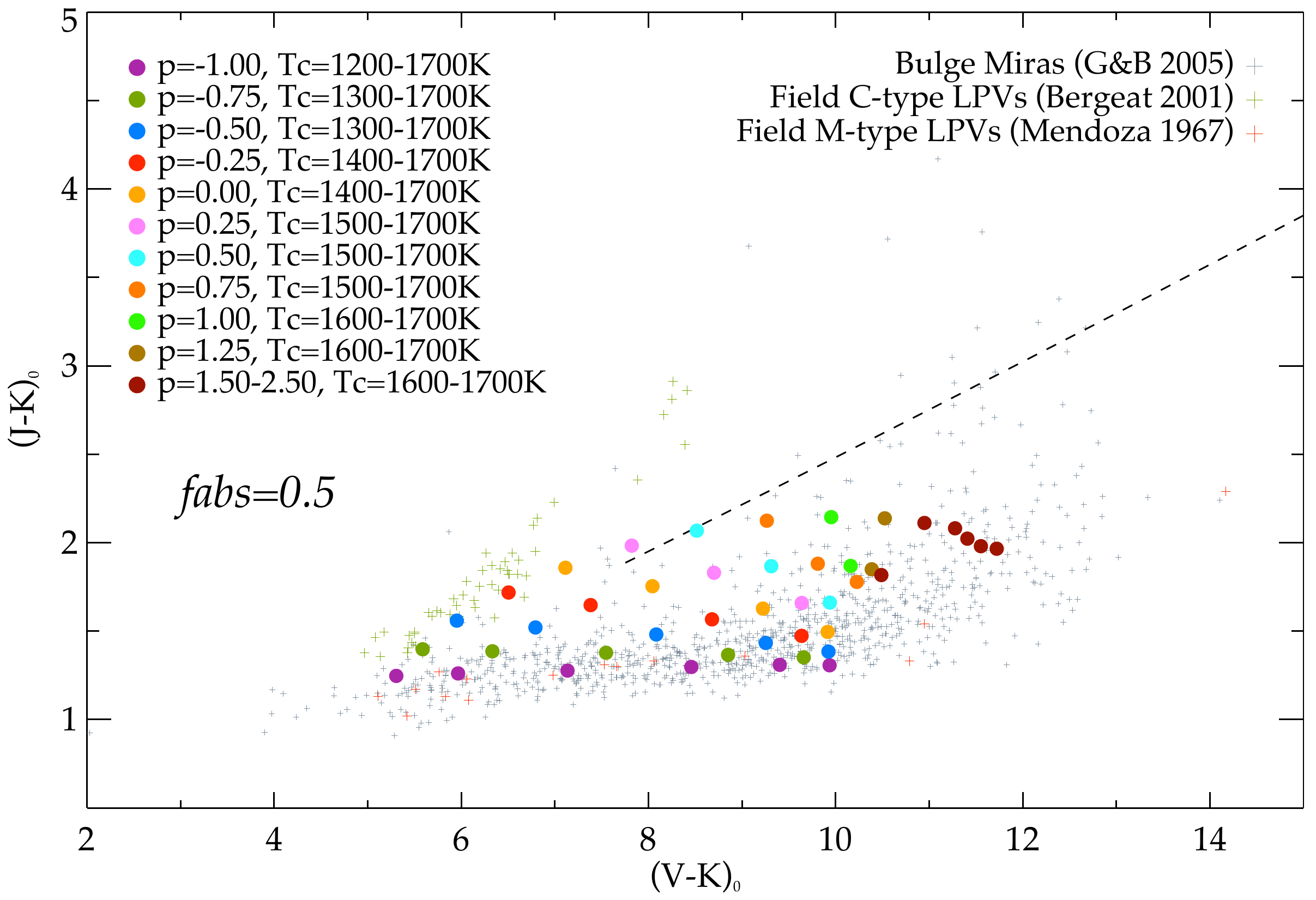}
  \caption{Phase-averaged photometry color-coded according to \mbox{$p$-value} for the models in set P. The higher the condensation temperature for a given \mbox{$p$-value} (circles with same color) the further the filled circles are from the bottom right corner. \textit{Top panel:} The fraction of true absorption is set to $f_{\mathrm{abs}}=1.0$. \textit{Bottom panel:} The fraction of true absorption is set to \mbox{$f_{\mathrm{abs}}=0.5$}. }
   \label{f_grid}
\end{figure}
The resulting colors of the models in set P, averaged over phase and color-coded according to $p$-value, are plotted in \mbox{Fig.~\ref{f_grid}}.  The averaged colors for the dynamical models in the grid with $f_{\mathrm{abs}}=1.0$ (top panel in \mbox{Fig.~\ref{f_grid}}) suggest that dust species with a large absorption cross-section in the near-IR, that is, with a wavelength-dependence corresponding to $p\gtrsim0.25$, are unlikely wind-drivers for M-type AGB stars. The resulting colors are too red, both in \mbox{($V$\,--\,$K$)} and ($J$\,--\,$K$), compared to the bulk of the observed values and in particular compared to the well-observed set of Miras in the top panel of Fig.~\ref{f_photoobs}. Similar criteria are valid for the grid with $f_{\mathrm{abs}}=0.5$ (bottom panel in \mbox{Fig.~\ref{f_grid}}), even if the resulting colors are not as red. A high condensation temperature also results in colors that are too red in ($J$\,--\,$K$) for both grids: the two highest 'arcs', corresponding to models with $T_{\mathrm{c}}=1600-1700$~K in Fig.~\ref{f_grid}, lie mostly outside the region of observed colors. 

Overall the photometric data covers different areas in the color-color plots for the two different values of $f_{\mathrm{abs}}$. This can be explained by the fact that more of the stellar radiation is thermally reprocessed by dust particles (i.e. absorbed at shorter wavelength and emitted at longer wavelengths) when the fraction of true absorption $f_{\mathrm{abs}}$ is higher, resulting in larger circumstellar reddening and higher values of \mbox{($J$\,--\,$K$)}. If we take mass-loss rates into account (see Table~\ref{t_dynmod} and the middle and bottom panel of Fig.~\ref{f_dynobs}) we see that for a given $p$, a higher mass-loss rate corresponds to a higher value of \mbox{($J$\,--\,$K$)}, although the trend is much weaker for grains with a lower fraction of true absorption. For  \mbox{($V$\,--\,$K$)}, however, there is no such simple correlation between color and mass-loss rate. It is also interesting to note that the high mass-loss models with $f_{\mathrm{abs}}=1.0$ (the upper 'arc' in the top panel of Fig.~\ref{f_grid}) are actually completely obscured and touch the line of black body emitters.

\subsection{Photometric variations}
\begin{figure}
\centering
\includegraphics[width=8.5cm]{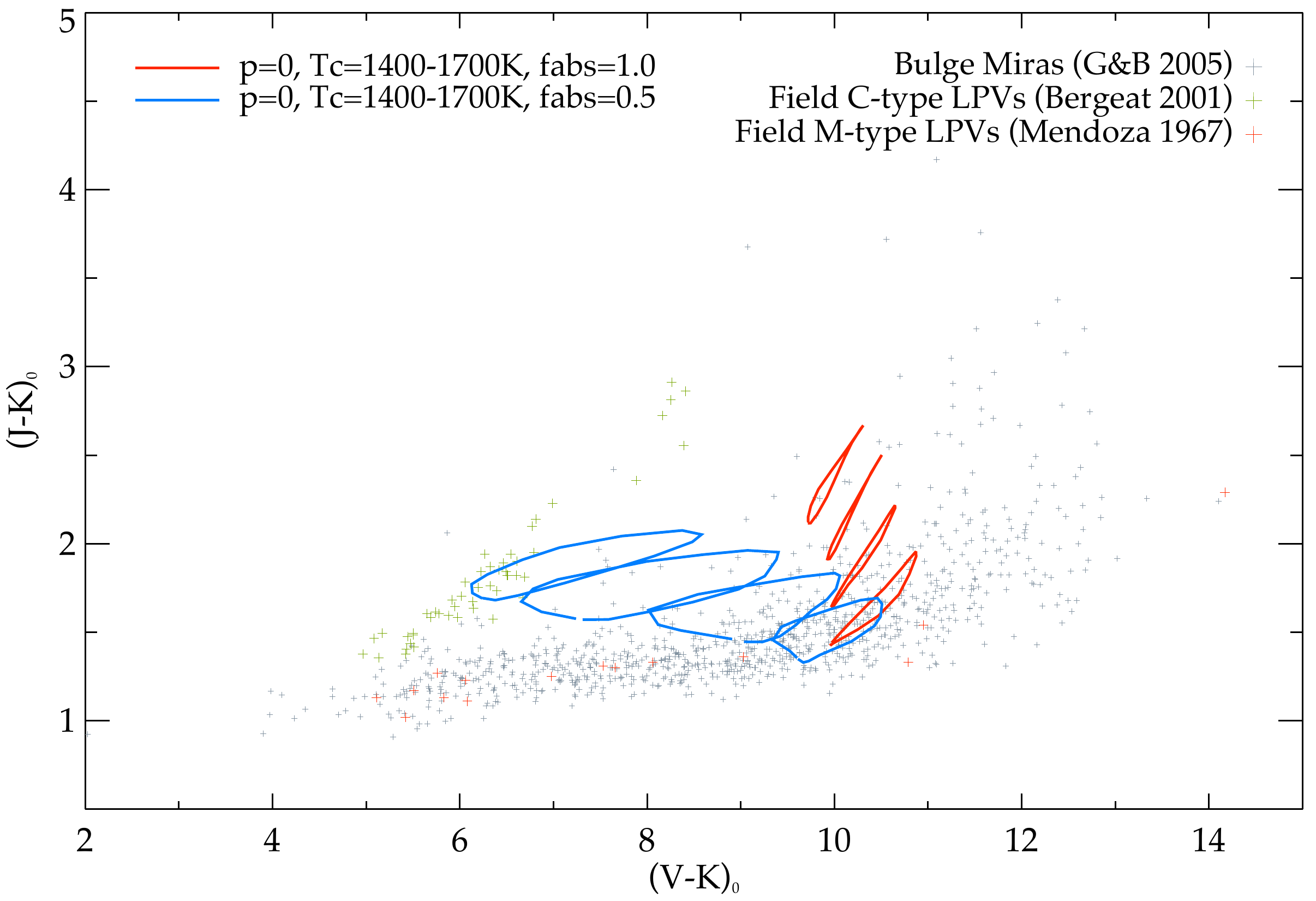}
\includegraphics[width=8.5cm]{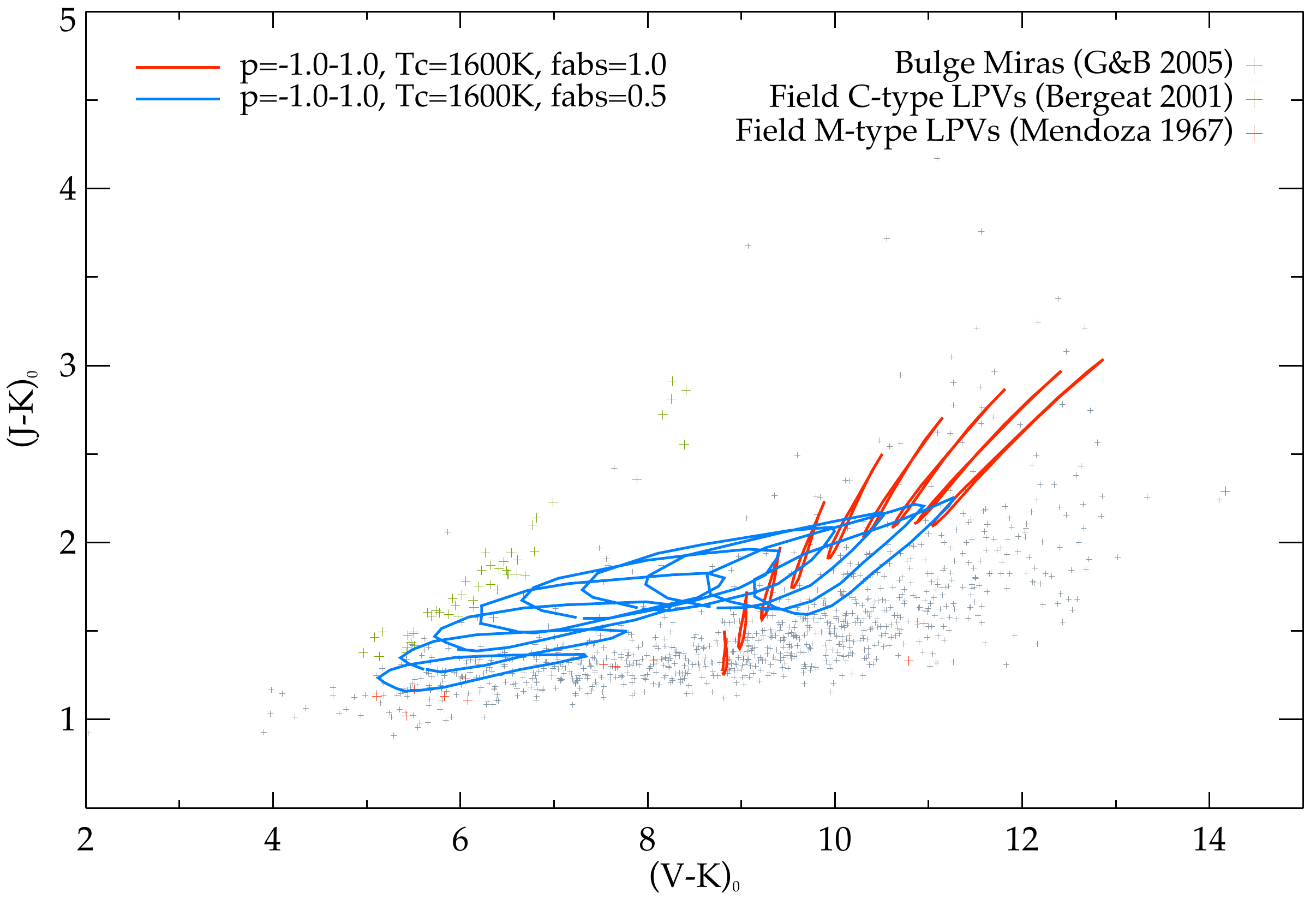}
  \caption{Photometric variations for models in set P, with \mbox{$f_{\mathrm{abs}}=1.0$} (red) and $f_{\mathrm{abs}}=0.5$ (blue). \textit{Top panel:} models with $p=0$ (grey dust) and $T_{\mathrm{c}}=1400-1700$\,K. The higher the condensation temperature the further the color loops are from the bottom right corner. \textit{Bottom panel:} models with $T_{\mathrm{c}}=1600$\,K and $p$ varying between $[-1.0,1.0]$. The higher the $p$-value the further the loops are from the bottom left corner. }
   \label{f_phgrey}
\end{figure}
In contrast to the phase-averaged colors discussed in the previous section, let us now study time-dependent photometry in detail. In the top panel of Fig.~\ref{f_phgrey} we plot the colors ($J$\,--\,$K$) and \mbox{($V$\,--\,$K$)} as a function of time for grey dust ($p=0$) and different condensation temperatures. The color variations plotted in red show photometric data for models with $f_{\mathrm{abs}}=1.0$ and the color variations plotted in blue show photometric data for models with $f_{\mathrm{abs}}=0.5$. For both sets of models, the color ($J$\,--\,$K$) increases with increasing condensation temperature. The reason for this is that an increased condensation temperature will move the dust formation zone closer to the stellar surface, where the density is higher. This leads to higher mass-loss rates, which will increase the circumstellar reddening and consequently produce higher \mbox{($J$\,--\,$K$)} values (see Paper I for further discussion). In the dynamical models with a lower fraction of true absorption, less radiation is thermally reprocessed.\footnote{Note that dust opacity $\kap$ in the equation of motion is the same for both grids, resulting in similar mass-loss rates.} Therefore the color \mbox{($J$\,--\,$K$)} will increase less with increasing mass-loss rate, as can be seen in the top panel of Fig.~\ref{f_phgrey}. 

\begin{figure}
\centering
\includegraphics[width=8.5cm]{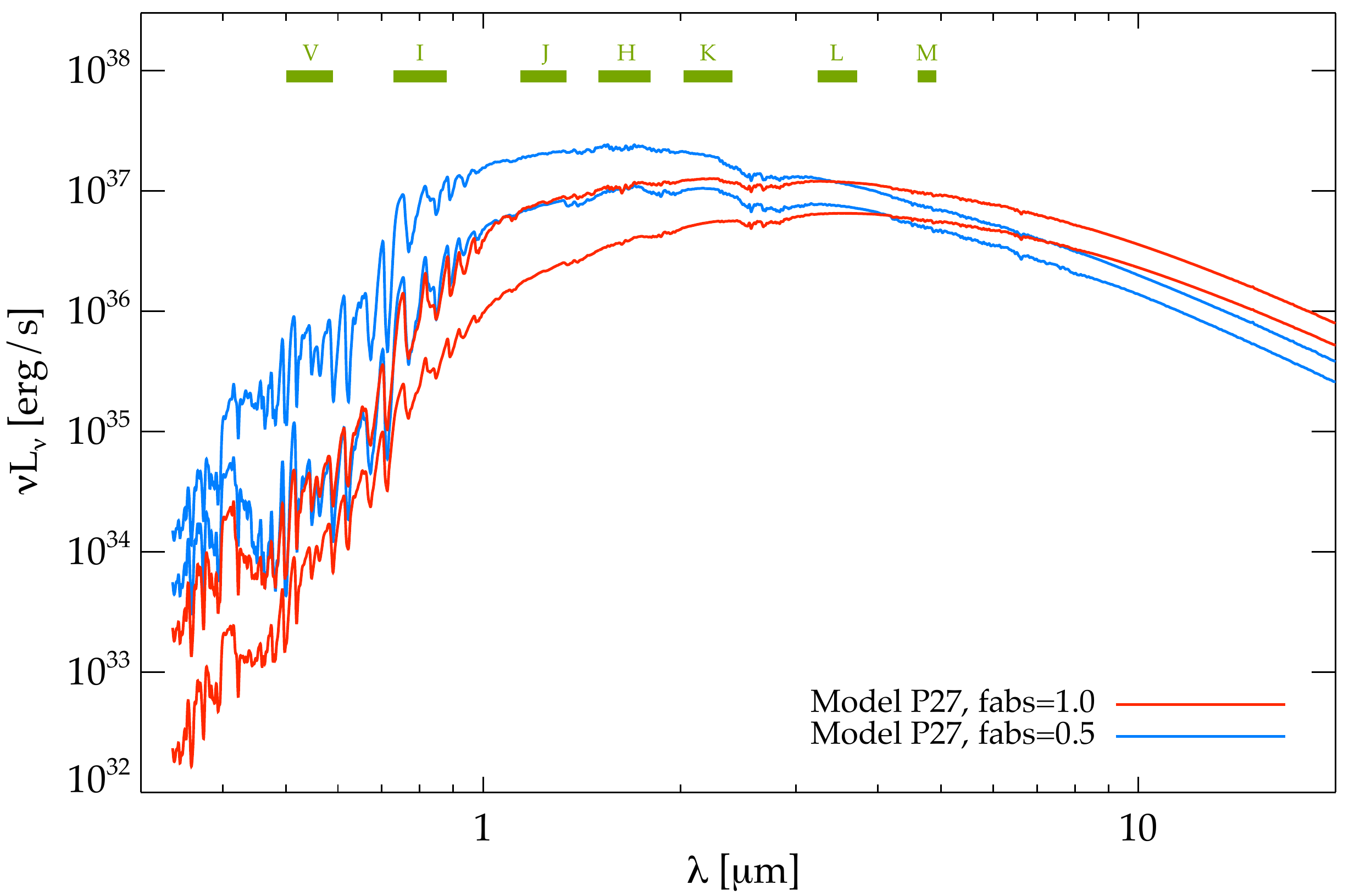}
  \caption{Spectra of the model P27  ($p=0$ and \mbox{$T_{\mathrm{c}}=1700$~K}), during a luminosity maximum (upper curve) and a luminosity minimum (lower curve), with $f_{\mathrm{abs}}=1.0$ (red) and $f_{\mathrm{abs}}=0.5$ (blue).}
   \label{f_spgrey}
\end{figure}

The understand why the color loops have such different shapes in models with $f_{\mathrm{abs}}=1.0$ compared to models with $f_{\mathrm{abs}}=0.5$ (red versus blue loops) we plot the spectra for two dynamical models with varying degrees of true absorption, but otherwise identical parameters. As can be seen in Fig.~\ref{f_spgrey}, the two spectra differ the most in the near-IR and visual regions, where the model with a lower fraction of true absorption emits significantly more stellar flux compared to the model with a higher fraction of true absorption. Another noticeable difference is that the visual magnitude varies more strongly with time when the fraction of true absorption is lower. This change in flux, caused by variations of the molecular features (mainly TiO) in the underlying atmosphere, becomes visible when dust is not obscuring the stellar radiation. Strong variations in the local gas temperature during a pulsation cycle drastically change the molecular abundances in these models, and consequently, how much of the visual flux is absorbed. These differences in the spectra are the reason why ($V$\,--\,$K$) is bluer and spanning a wider range in the color-color diagram for models with $f_{\mathrm{abs}}=0.5$ compared to models with $f_{\mathrm{abs}}=1.0$. The circumstellar reddening is also less pronounced in the model with $f_{\mathrm{abs}}=0.5$, which results in  smaller flux variations in the $J$ band, and consequently, smaller variations in ($J$\,--\,$K$).

\begin{figure}
\centering
\includegraphics[width=8.5cm]{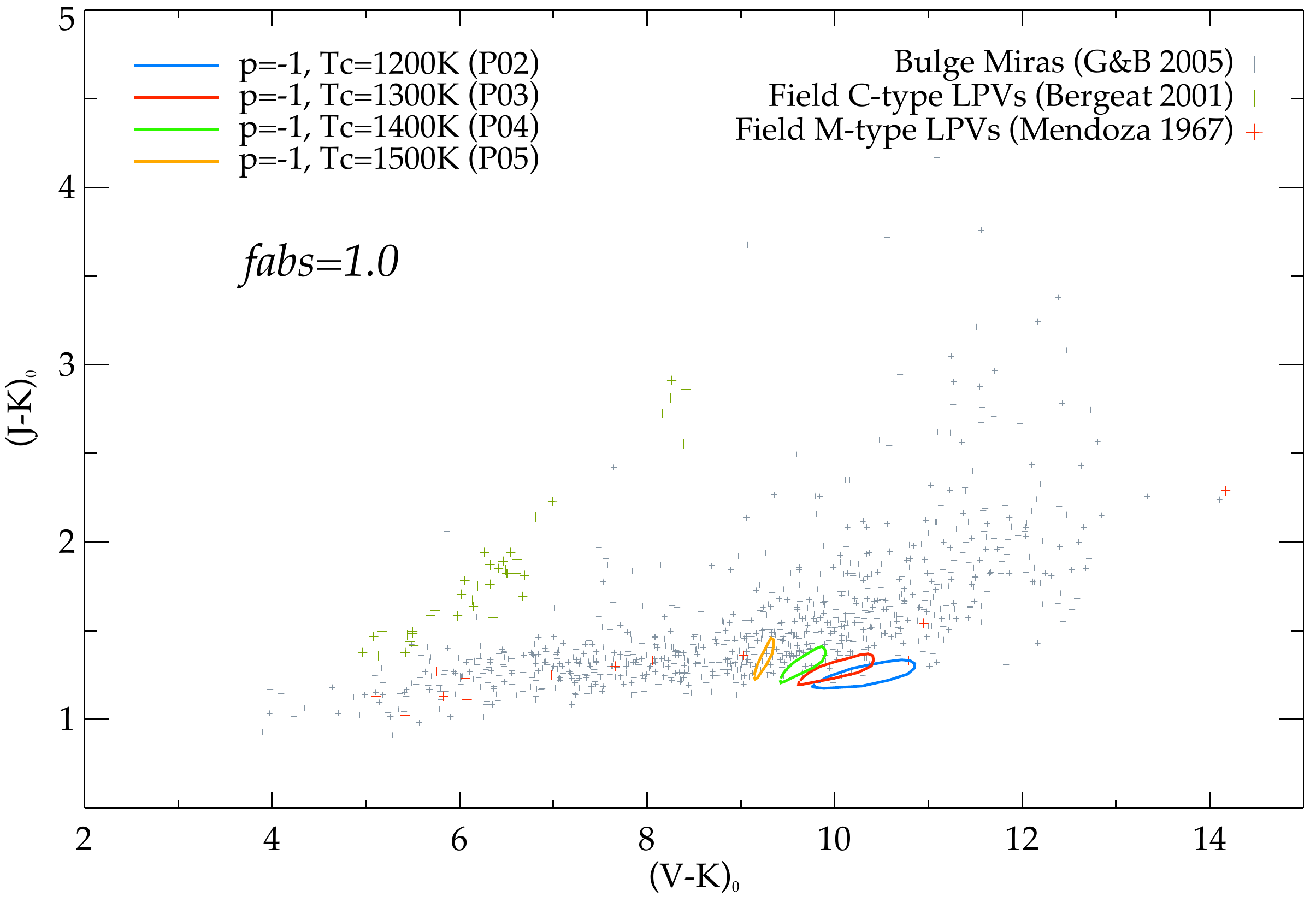}
\includegraphics[width=8.5cm]{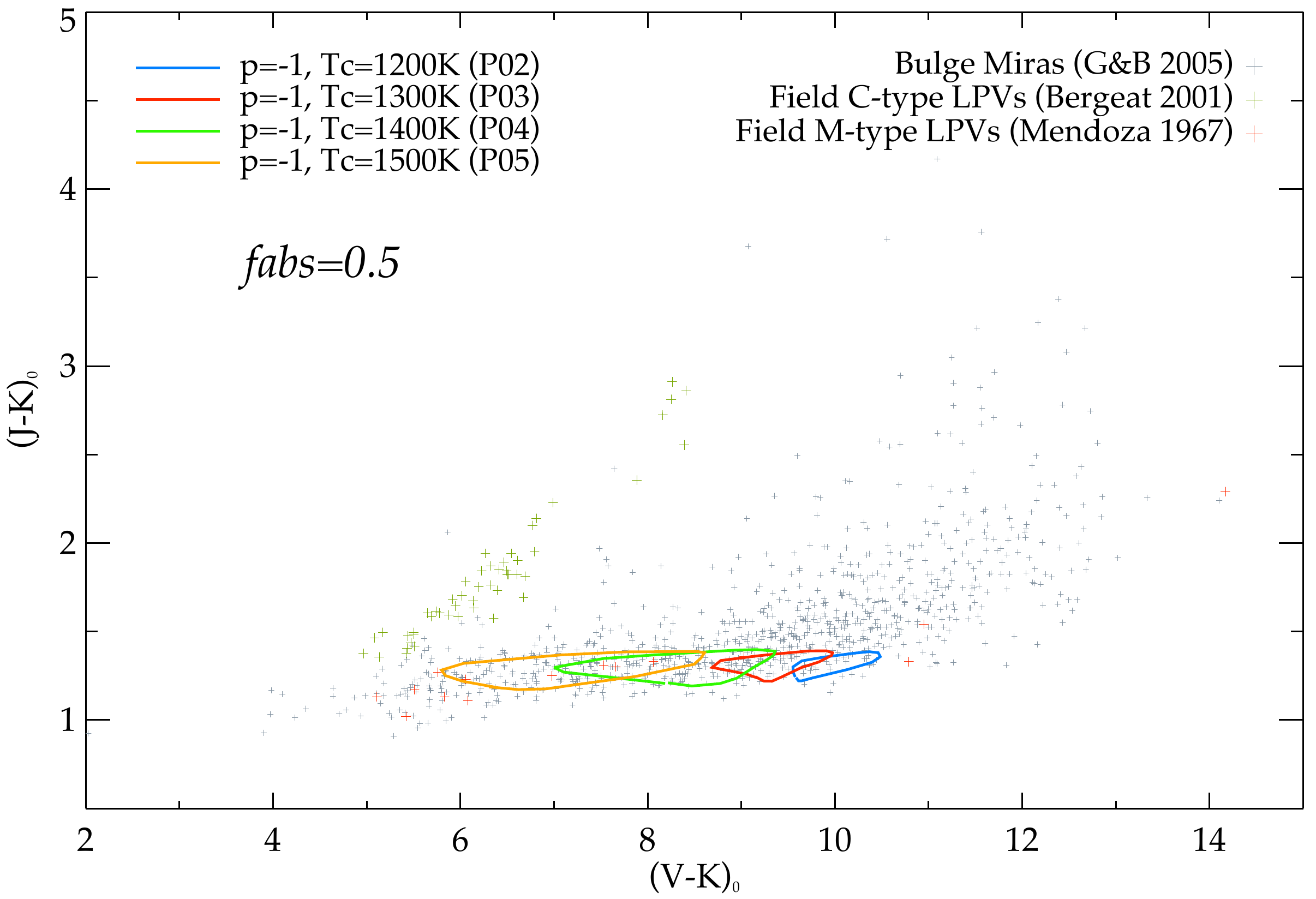}
  \caption{Photometry variations for models P02-P05 in set P (\mbox{$p=-1$} and \mbox{$T_{\mathrm{c}}=1200-1500$~K}).  \textit{Top panel:} The fraction of true absorption is set to \mbox{$f_{\mathrm{abs}}=1.0$.} \textit{Bottom panel:} The fraction of true absorption is set to \mbox{$f_{\mathrm{abs}}=0.5$}.}
   \label{f_phpn100}
\end{figure}

To systematically study how the photometric variations change with different values of $p$ (i.e. different wavelength dependences of the dust opacity), we plot ($J$\,--\,$K$) and ($V$\,--\,$K$) for a set of dynamical models where $p$ is varied between $[-1,1]$ and the condensation temperature is kept fixed at $T_{\mathrm{c}}=1600$\,K \mbox{(see the bottom panel of Fig.~\ref{f_phgrey})}.  If we compare the result with the observational data in Fig.~\ref{f_photoobs}, it is clear that the photometric data from models with $f_{\mathrm{abs}}=1.0$ do not reproduce the time-dependence of the observed colors at all, nor are the majority of them situated within the region of observed AGB stars. Photometric data from models with $f_{\mathrm{abs}}=0.5$, on the other hand, have loops very similar to the observed colors when $p\lesssim 0.25$, with small variations in ($J$\,--\,$K$) and a large span in ($V$\,--\,$K$). For larger $p$-values the color ($J$\,--\,$K$) varies too much compared to the colors fitted from observations in Fig.~\ref{f_photoobs}. 

As mentioned in Sect.~\ref{s_phdetdus}, the resulting colors from the dynamical models in set D, using a detailed description of the growth of Mg$_2$SiO$_4$ grains, are consistent with the colors fitted from observations. This grain material has a very low absorption cross-section in the visual and near-IR wavelength region and a power law index of $p\approx-1$ (see Fig.~\ref{f_pl}). In an attempt to duplicate how the colors vary with time for the models in set D, we plot ($J$\,--\,$K$) and ($V$\,--\,$K$) for dynamical models using a parameterized dust opacity with $p=-1$, $T_{\mathrm{c}}=1200-1500$\,K and $f_{\mathrm{abs}}=0.5$ (models P02-P05 in set P). The resulting photometric variations are shown in the bottom panel of Fig.~\ref{f_phpn100}. For comparison we also plot the corresponding variations with $f_{\mathrm{abs}}=1.0$ in the top panel of Fig.~\ref{f_phpn100}. Indeed, for $f_{\mathrm{abs}}=0.5$ the colors vary with time similar to how the colors vary for the dynamical models in set D and they are situated within the same region of the color-color diagram (see bottom and middle panel of Fig.~\ref{f_photoobs}). However, the span of ($V$\,--\,$K$) for each model is more restricted. One possible explanation for this discrepancy is the question of applicability of the power-law fit in the visual region. The real optical data of Mg$_2$SiO$_4$ grains and the power law fit diverge quickly outside the wavelength range used for the fit, as can be seen in Fig.~\ref{f_pl}. This could affect the magnitude in the visual band. Another possible explanation is that the Mg$_2$SiO$_4$ grains have such low absorption cross-sections that we would need to use a lower value for $f_{\mathrm{abs}}$ to reproduce the span of \mbox{($V$\,--\,$K$)}. 

\subsection{Conclusions from set P}
\label{s_concl1}
To sum up the discussion so far, it is evident that the chemical and optical properties of the wind-driving dust species affect the resulting spectra and photometry strongly and that the realistic color variations produced by the models in set D are not a trivial result. By comparing observed visual and near-IR colors from M-type AGB stars to photometry from the models in set P, covering a range of chemical and optical dust properties, we can conclude that in order to produce colors with similar time-dependence as indicated by observations, i.e., small  variations in ($J$\,--\,$K$) and spanning a larger range in ($V$\,--\,$K$), the dust material needs to be quite transparent in this wavelength region. The large variation in \mbox{($V$\,--\,$K$)}, caused by molecular species such as TiO, can only be seen if the circumstellar envelope is not optically thick. In addition, transparent grain materials will lead to weaker circumstellar reddening, and thereby, smaller variations in ($J$\,--\,$K$).

Furthermore, considering the location of the models in the color-color diagram, the absorption coefficient has to be close to grey or increase with increasing wavelength ($p\lesssim 0.25$) and the condensation temperature has to be equal to or below 1500\,K, in order not to cause too red values in ($V$\,--\,$K$) and ($J$\,--\,$K$), respectively. There is a possibility for a hypothetical dust material that has an absorption coefficient that increases steeply with increasing wavelength ($p\lesssim-0.75$) and a high condensation temperature ($T_{\mathrm{c}}>1500$\,K) to produce colors that are not too red in ($V$\,--\,$K$) or ($J$\,--\,$K$), but among the optical data currently available for dust species which are assumed to exists in M-type AGB stars, no such dust species can be found (see Table~\ref{t_more}). 

\section{Evaluating suggested wind-drivers}
\label{s_spg}
\begin{table}
\caption{Examples of $p$ and $T_{\mathrm{c}}$ values for dust species likely to exist in the circumstellar environments of AGB stars.}             
\label{t_more}      
\centering                          
\begin{tabular}{l l r l c l}        
\hline\hline                 
\# & Dust species & $p$ & Ref. &$T_{\mathrm{c}}$ [K]  &Ref. \bigstrut[t] \bigstrut[b]\\    
\hline  
  1 & Fe & 2.4 & 3 &1050 &1 \bigstrut[t] \\       
  2 & FeO & 1.6 & 4 & 900 & 1\\
  3 & Mg$_{0.5}$Fe$_{0.5}$O & 1.9 & 4 & 1000 & 1\\
  4 & Mg$_{0.5}$Fe$_{0.5}$SiO$_3$ & 2.4 & 4 & 1100 & 1\\	
  5 & Fe$_2$SiO$_4$ & 1.7 & 5 &1050 & 1\\
  6 & MgFeSiO$_4$ & 2.3 & 6 & 1100 & 1 \\ 
  7 & MgSiO$_3$ & $-0.5$ & 7 & 1100 & 1\\  
  8 & Mg$_2$SiO$_4$ & $-0.9$ & 7 & 1100 & 1\\   
  9 & SiO$_2$ & 0.2 & 8 & 1050 & 1 \\ 
  10 & Al$_2$O$_3$ & 1.7 & 9,10 & 1400 & 1\\
  11 & MgAl$_2$O$_4$ & $-1.2$ & 5 & 1150 & 1 \\
  12 &TiO$_2$ & 0.1 & 5 &  1100 & 2  \bigstrut[b] \\
\hline   
  13 & amC &1.2 & 11 &1700 & 2  \bigstrut[t] \\  
  14 & SiC & $0.3$ & 12 & 1100 & 1 \bigstrut[b] \\
\hline   
\end{tabular}
\tablefoot{The dust species above the separating line are likely to exist in the circumstellar environment of M-type AGB stars, whereas the dust species below are found in C-type AGB stars. Note that the $p$-value for SiO$_2$ is from interpolated optical data and that the optical data in the near-IR for Al$_2$O$_3$ is uncertain (see discussion in Sect. 6 of Paper I).}
\tablebib{(1)~\citet{gailminerology};
(2) \citet{latt78}; (3) \citet{ord88fe}; (4) \citet{hen95}; (5) \citet{zeid11}; 
(6) \citet{dor95}; (7) \citet{jag03}; (8) \citet{pal85};
(9) \citet{koi95cor}; (10) \citet{beg97}; (11) \citet{roul91amc}; (12) \citet{peg88sic}.
}
\end{table}

In Paper I we looked at dynamical criteria, i.e. what combination of optical and chemical properties are needed for a dust species to form close enough to the star to initiate mass outflows, when evaluating different dust materials in search for possible wind-drivers in M-type AGB stars. In this section we evaluate specific dust material from a photometric perspective instead, to gain further insight into the dust properties that are needed to both trigger winds and produce spectra and photometry consistent with observations.

Dust species that include Fe and for which optical data in the near-IR exist in the literature have $p$-values outside the parameter space which gives realistic colors (e.g. \mbox{\#1-6} in Table~\ref{t_more}). Among the Fe-free silicates and oxides, both Mg$_2$SiO$_4$ and MgSiO$_3$ fulfill the constraints. The chemical and optical properties of TiO$_2$ and MgAl$_2$O$_4$ also fall within the parameter space, but the constituent elements have too low abundances, leading to insufficient radiative acceleration. Uncertainties concerning the optical data for SiO$_2$ and Al$_2$O$_3$ in the near-IR wavelength region makes them hard to evaluate (a more detailed discussion on these issues can be found in Paper I) even though Al$_2$O$_3$ is not a probable candidate due to low element abundance of Al. SiO$_2$, on the other hand, consists both of abundant atomic elements and is very transparent in the near-IR (i.e. thermally stable close to the star) and could be of interest for AGB stars where the C/O-ratio approaches unity and most of the oxygen is bound in CO-molecules, making it less available for dust formation. 

\subsection{Carbon dust as wind-driver in M-type AGB stars?}
\begin{figure}
\includegraphics[width=8.5cm]{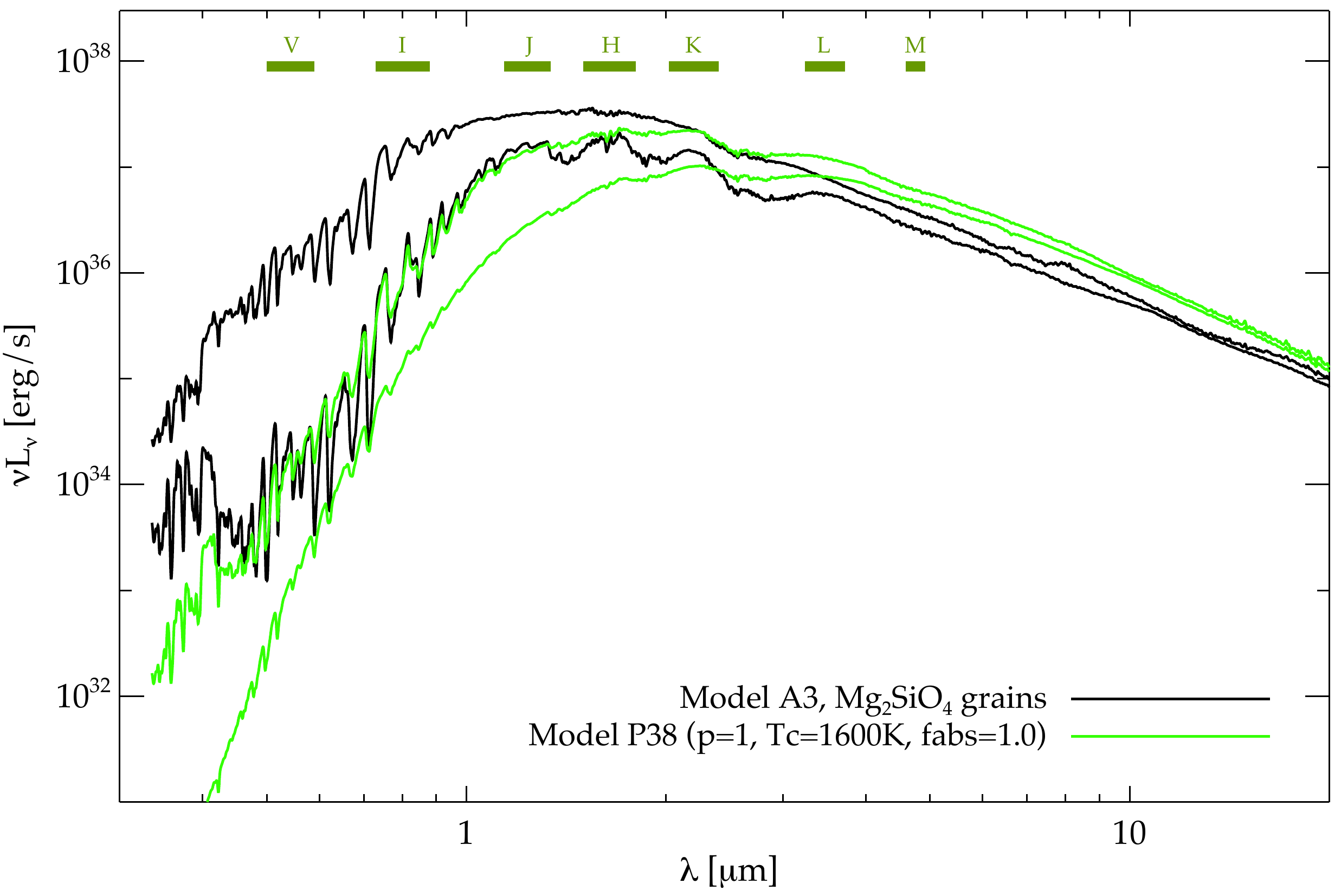}
\includegraphics[width=8.5cm]{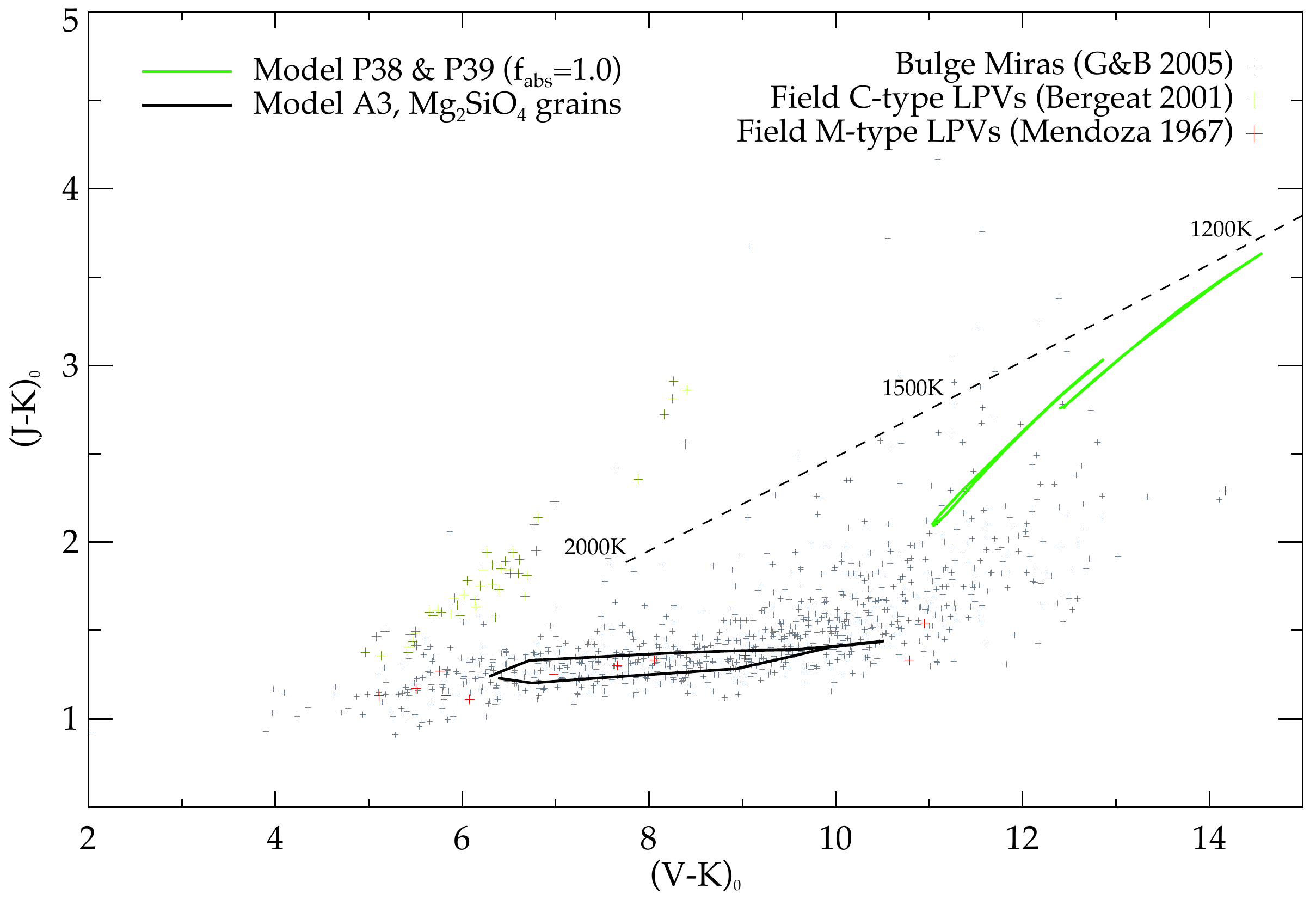}
\caption{Spectra and photometry for wind models using a parameterized dust description simulating carbon grains.  \textit{Top panel:} Spectra for the model P38 with $f_{\mathrm{abs}}=1.0$ (green), and for comparison, the model A3 (black), during luminosity minimum and maximum. \textit{Bottom panel:} Photometric variations for the models P38-P39 ($f_{\mathrm{abs}}=1.0$) and A3. Over-plotted are the locations of simulated blackbody emitters in the range of $T_\star = 1100-2000$\,K (dashed line).}
\label{f_spcarb}
\end{figure}

As a consequence of the results by \cite{woi06fe}, which indicated problems with the dust-driven wind schemes based on Fe-bearing silicates, an alternative scenario for the mass loss of M-type AGB stars was suggested by \cite{hof07carb}. They speculated that the outflows from these stars might be driven by a small amount of carbon grains, with Fe-free silicate grains forming as a by-product in the wind and producing the observed mid-IR features. In chemical equilibrium, most of the carbon in an atmosphere with C/O\,$<$\,1 will be bound in the stable CO-molecule, and will therefore not be available for dust formation. However, deviations from equilibrium gas phase chemistry due to strong atmospheric shock waves are not improbable \citep[e.g.][]{cher06,cher11}. The question remains if carbonaceous grains can actually form from freed carbon atoms in the wake of the shock waves, but nevertheless we here investigate the effect on spectra if outflows in M-type AGB stars indeed were driven by carbon grains. 

The absorption coefficient of amorphous carbon can be fitted well by a power-law function, even quite a bit outside the wavelength range where the star radiates most of its flux \mbox{(see Fig.~\ref{f_pl}}), with a value of $p\approx 1$. Given the high absorption cross-section in the near-IR of amorphous carbon, the momentum transfer from stellar photons to the carbon grains will be dominated by true absorption and not scattering. To investigate the effects of carbon grains in a M-type AGB star we therefore study the spectrum of model P38 in set P, with $p=1$, $T_{\mathrm{c}}=1600$~K and $f_{\mathrm{abs}}=1.0$. For comparison, we also consider the spectrum of the model A3 in set D. This latter model was chosen since the photometry agrees very well with the colors fitted from observations (see Fig.~\ref{f_photoobs}). 

The synthetic spectra at luminosity extremes are shown in the top panel of Fig.~\ref{f_spcarb}. It is obvious that a grain material with optical properties similar to amorphous carbon will greatly suppress the molecular features in the visual and near-IR, especially during the luminosity minimum, and cause significant circumstellar reddening. This is something that is not observed in the spectra of M-type AGB stars. In contrast, these stars show strong molecular features in the visual and less circumstellar reddening compared to carbon stars. On the other hand, Fig.~\ref{f_spcarb} also shows that dynamical models which include a detailed description of Mg$_2$SiO$_4$ grains produce spectra with the strong molecular features in the visual intact and low circumstellar reddening.

In addition to the effect on the spectra in the visual wavelength region, the resulting photometry from model P38 does not agree with the observed values for M-type AGB stars, as can be seen in the bottom panel of Fig.~\ref{f_spcarb}. Both ($V$\,--\,$K$) and ($J$\,--\,$K$) are too red compared to values by \cite{men67} or \citet{gbmiras} and the photometric variations differ significantly from the photometric variations derived from observations (see top panel in Fig.~\ref{f_photoobs}). Given these results, it is clear that the amount of carbon grains necessary to produce an outflow in M-type AGB stars would affect the spectra in ways that are not compatible with observations and that the scenario presented in \cite{hof07carb} is not viable.

\section{Summary and conclusions}
\label{s_concl2}

It has long been speculated that winds of M-type AGB stars are driven by radiative acceleration on silicate grains observed in the circumstellar envelopes of these stars. Recent theoretical results by \cite{woi06fe}, showing that silicate grains have to be virtually Fe-free in the close vicinity of AGB stars, however, raised doubts about this scenario. The low near-IR absorption cross-sections of such Fe-free grains are not sufficient to trigger outflows. As a response to this result \citet{hof08bg} argued that wind of M-type AGB stars may be driven by photon scattering on Fe-free silicates, provided that the grains grow to sizes of about $0.1-1\mu$m. Strong observational support for this scenario was recently given by \cite{norr12}, who detected dust particles of sizes $\sim0.3\,\mu$m in the close circumstellar environment of three M-type AGB stars, using multi-wavelength aperture-masking polarimetric interferometry. 

In this paper we provide further support for the idea of winds driven by photon scattering on dust by presenting photometry for the set of self-consistent wind models in \citet{hof08bg}.\footnote{By self-consistent we here mean radiation-hydrodynamical models that include a time-dependent description for the grain growth, grain-size dependent dust opacities and frequency-dependent radiative transfer. Note also that the \textit{a posteriori} radiative transfer is done consistently with the dynamical computations, only differing in the number of frequency points used.} The resulting $V$, $J$ and $K$ photometry reproduces remarkably well both the values and the time-dependent behavior, i.e. small variations in ($J$\,--\,$K$) and spanning a larger range in ($V$\,--\,$K$), of photometric observations of M-type AGB stars. To our knowledge, these are the first self-consistent wind models for M-type AGB stars that reproduce well both observed dynamical properties, such as wind velocities and mass-loss rates (see Fig.~\ref{f_dynobs}), and photometry in the visual and near-IR wavelength region (see Fig.~\ref{f_photoobs}). 

To determine if the trends in photometry are a generic property of the dynamical models or if they constrain the optical and chemical properties of the grain material driving the wind, we explore photometry for a set of models presented in \cite{bladh12}. These models use a parameterized dust description and span a range of optical and chemical dust properties, including varying transparency. Looking at the photometry for this set, it is evident that the chemical and optical properties of the wind-driving dust species affect the resulting spectra and photometry strongly, and that the realistic color variations produced by the models using a detailed dust description for the growth of Mg$_2$SiO$_4$ grains are not a trivial result. The wind-driving dust species need to have a low absorption cross-section in the visual and near-IR to reproduce the typical time-dependent behavior observed in M-type AGB stars. The large variation in ($V$\,--\,$K$), mainly due to molecular species such as TiO, can only be seen if there is no substantial absorption due to dust in the circumstellar envelope. In addition, transparent grain materials will lead to less thermal reprocessing of the stellar radiation, resulting in small variations in ($J$\,--\,$K$). The photometric observations also place constraints on the optical and chemical properties of the wind-driving dust species ($p\lesssim 0.25$ and $T_{\mathrm{c}}\lesssim 1500$~K) in order to avoid large values in ($V$\,--\,$K$) and ($J$\,--\,$K$). 

Concerning specific grain materials, we conclude that carbon grains \citep[suggested by][]{hof07carb} are not viable as wind-drivers in M-type AGB stars since they would affect the spectra in ways that are not compatible with observations. Strong candidates for wind-driving dust species in M-type AGB stars are Fe-free silicates, such as Mg$_2$SiO$_4$ and MgSiO$_3$, provided that they reach grain sizes where scattering is efficient. They consist both of abundant materials and are quite transparent in the visual and near-IR. Another possibility could be SiO$_2$, especially for stars with a C/O-ratio close to unity where most of the oxygen is bound in CO-molecules, making it less available for dust formation.

\begin{acknowledgements}
Sincere thanks are given to M.A.T.~Groenewegen and J.A.D.L.~Blommaert who provided mean OGLE-$V$-magnitudes for the Galactic Bulge Miras and to B.~Gustafsson for valuable feedback and insights. 
      
We acknowledge with thanks the variable star observations from the AAVSO International Database, contributed by observers worldwide and used in this research. This research has made use of (i) NASA's Astrophysics Data System, and (ii) the NASA/IPAC Infrared Science Archive, which is operated by the Jet Propulsion Laboratory, California Institute of Technology, under contract with the National Aeronautics and Space Administration. The computations were performed on resources provided by the Swedish National Infrastructure for Computing (SNIC) at UPPMAX. 

This work has been funded by the Swedish Research Council (\textit{Vetenskapsr\aa det}) and the Austrian Science Fund (FWF): P21988-N16. 
BA acknowledges support from Austrian Science Fund (FWF): AP23006 \& AP23586 and from contract ASI-INAF I/009/10/0. 

\end{acknowledgements}

\bibliographystyle{aa}
\citeindextrue
\bibliography{references}

\end{document}